\documentclass[pra, twocolumn,aps,superscriptaddress,showpacs]{revtex4-2}
\usepackage{subfigure}
\usepackage{psfrag}
\usepackage{dcolumn}
\usepackage{bm}
\usepackage{color}
\usepackage{latexsym}
\usepackage{epstopdf}
\usepackage[english]{babel}
\usepackage{times}
\usepackage{latexsym}
\usepackage{natbib}
\usepackage{graphicx}
\usepackage{epsf}
\usepackage{subfigure}
\usepackage{amsmath}
\usepackage{mathtools}
\usepackage{amssymb}
\usepackage{amsfonts}
\usepackage{bm}
\usepackage{appendix}
\usepackage{lipsum, babel}
\usepackage{soul}
\usepackage{hyperref}
\hypersetup{colorlinks=true,citecolor=myblue,linkcolor=blue,urlcolor=myblue}
\DeclareUnicodeCharacter{00A0}{ }

\newcommand{\vt}{v_{\mathrm{th}}}

\newcommand{\kp}{\vec{k}_\mathrm{p}}
\newcommand{\kc}{\vec{k}_\mathrm{c}}

\newcommand{\OD}{\text{OD}}
\newcommand{\ODeit}{\text{OD}_\text{EIT}}
\newcommand{\geit}{\gamma_\text{EIT}}
\newcommand{\aeit}{\alpha_\mathrm{p}(\delta=0)}
\newcommand{\keff}{\vec{k}_\text{eff}}

\definecolor{mygrey}{gray}{0.35}
\definecolor{myblue}{rgb}{0.2,0.2,0.8}
\definecolor{myzard}{cmyk}{0,0,0.05,0}
\definecolor{mywhite}{rgb}{1,1,1}
\definecolor{myred}{rgb}{1,0.,0.3}
\definecolor{myblack}{rgb}{0,0,0}

\usepackage{dsfont}

\begin{document}

\widetext


\title{A practical guide to  electromagnetically induced transparency in atomic vapor}
\author{Ran Finkelstein}
\affiliation{Department of Physics of Complex Systems, Weizmann Institute of Science, Rehovot 7610001, Israel}
\affiliation{Division of Physics, Mathematics and Astronomy, California Institute of Technology, Pasadena, CA 91125, USA}
\author{Samir Bali}
\affiliation{Department of Physics, Miami University, Oxford, Ohio 45056, USA}
\author{Ofer Firstenberg}
\affiliation{Department of Physics of Complex Systems, Weizmann Institute of Science, Rehovot 7610001, Israel}
\author{Irina Novikova}
\affiliation{Department of Physics, William and Mary, Williamsburg, Virginia 23187, USA}
\date{\today}
\begin{abstract}
This tutorial introduces the theoretical and experimental basics of Electromagnetically Induced Transparency (EIT) in thermal alkali vapors. We first give a brief phenomenological description of EIT in simple three-level systems of stationary atoms and derive analytical expressions for optical absorption and dispersion under EIT conditions. Then we focus on how the thermal motion of atoms affects various parameters of the EIT system. Specifically, we analyze the Doppler broadening of optical transitions, ballistic versus diffusive atomic motion in a limited-volume interaction region, and collisional depopulation and decoherence. Finally, we discuss the common trade-offs important for optimizing an EIT experiment and give a brief "walk-through" of a typical EIT experimental setup. We conclude with a brief overview of current and potential EIT applications. 
\end{abstract}

\maketitle
\section{Introduction}

\noindent Electromagnetically induced transparency (EIT)~\cite{harris'97pt,FleischhauerRevModPhys05} may be described almost like a magic tool: it makes opaque objects transparent and slows down light pulses to a crawl, even trapping these light pulses inside material objects. These make it a fun topic for public lectures, but more importantly in the last decades it has inspired a myriad of new applications, from precise atomic clocks and magnetometers~\cite{vanier05apb,kitchingAPR2018} to quantum information tools~\cite{BeausoleilJMO2004,Firstenberg_2016,Ma_2017,wei2022realworld}. 
In terms of its importance for atomic, molecular, and optical physics, the introduction of the concept of EIT may be compared with the demonstration of optical pumping several decades earlier~\cite{happer72} . Optical pumping gave physicists the ability to control atomic populations by means of optical fields. Similarly, EIT extended this control into the realm of coherent superpositions and quantum states and, in parallel, allows for manipulation of light by atoms, ushering a new approach for realization of strong light-atom coupling using collective enhancement~\cite{scully'92physrep,fleishhauerLukinPRL00,lukin03rmp}. Unlike the standard QED single-atom approach that requires strong coupling of an individual atom to a photonic mode, here the strong coupling is due to the collective enhancement provided by the large ensemble of identical atoms. No high-quality cavity is required, and the resulting collective atomic state is very robust and can faithfully preserve quantum information originally carried by the optical probe. 

\begin{figure}[b]
    \centering
    \includegraphics[width=0.95\columnwidth]{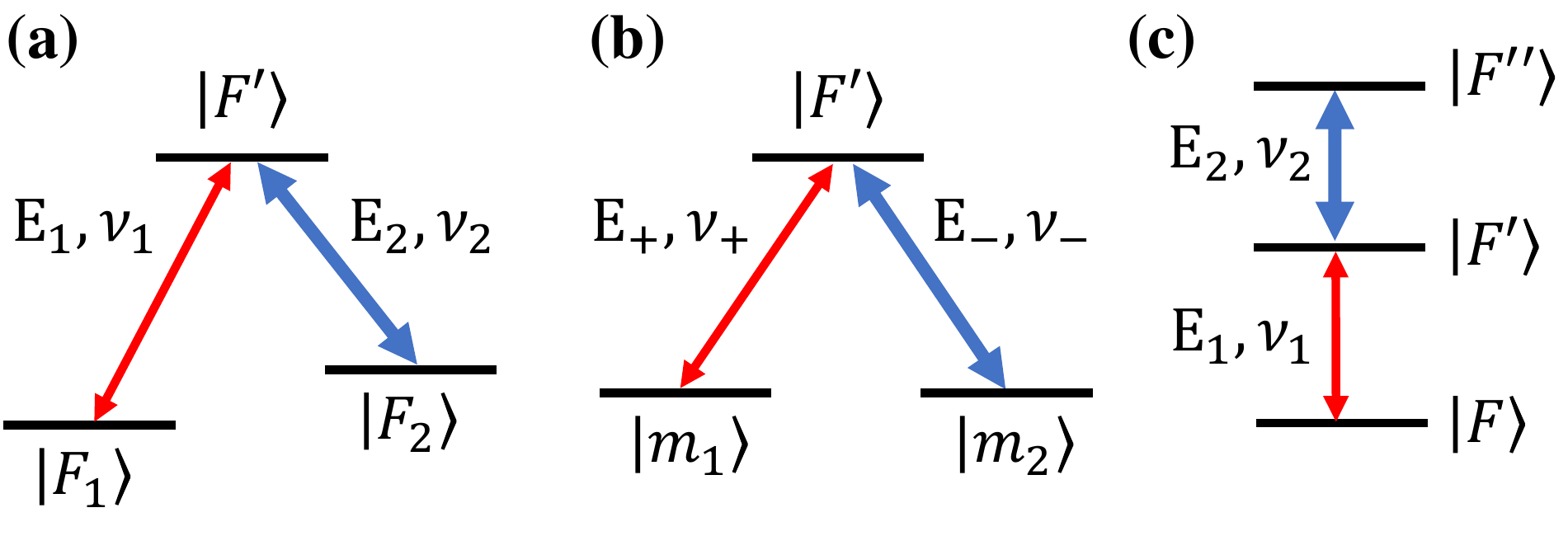}
    \caption{\textbf{The most common EIT configurations.} (a) Hyperfine $\Lambda$ system, in which two optical fields connect two hyperfine sublevels of an electronic ground state with a common excited state. (b) Zeeman $\Lambda$ system, in which two optical fields of orthogonal (usually circular) polarizations, and often with the same frequency, connect two Zeeman sublevels of the same hyperfine state to a common excited state. (c) Ladder system, in which two optical fields establish coherence between a ground and a highly excited (\textit{e.g.}, Rydberg) electronic state. }
    \label{fig:EITconfigs}
\end{figure}

The essence of the EIT effect is the strong coupling of a (usually weak) optical probe field to a long-lived excitation of an emitter by means of another, strong control field. The resulting two-photon transitions give optical access to high-Q quantum superpositions while avoiding optical losses associated with individual optical transitions. The term ``electromagnetically induced transparency'' became widely used after its introduction in 1991~\cite{HarrisPhysRevLett.66.2593}, although the essential concept was first described a few years earlier~\cite{kocharovskayaJETP86}, and its more humble incarnation -- coherent population trapping (CPT) -- was known since 1975 and had been explored for metrological applications~\cite{arimondo'76,arimondo'96po}. By now, EIT has been studied in a broad range of physical systems: atoms, molecules, 
ions, plasmas, crystals, plasmons, etc. In atoms, EIT typically utilizes two-photon transitions between two spin states of the same electronic level or  between ground and highly excited electronic states in, correspondingly, $\Lambda$ and ladder configurations, shown in Fig.~\ref{fig:EITconfigs}~\footnote{In principle, EIT can be also observed in a $V$ scheme, in which two excited states are connected to a common ground state. However, this configuration does not offer much advantage for atomic systems~\cite{PhysRevA.52.2302} and is typically explored for other platforms~\cite{PhysRevA.83.063419,PhysRevB.79.115420}}. While narrow transparency resonances can be observed in both cold and hot atoms, thermal motion in atomic vapor adds a new dimension to the EIT treatment and results in additional restrictions or opportunities, depending on the situation. 

In this paper we focus on properties of EIT in thermal vapors of alkali metals. One of the key advantages of such a platform is its simplicity: while a typical cold atom apparatus requires a vacuum system and additional infrastructure for cooling and trapping atoms, atomic vapor can be contained in an evacuated transparent cell with a simple heater to control atomic vapor density. This simplicity gives vapor cell-based EIT experiments great flexibility to adapt to the requirements of different applications. For example, chip-scale vapor cells and vapor-filled fibers and waveguides enable development of compact EIT-based tools and photonic elements~\cite{zektzer_atomphoton_2021}. It is also usually fairly straightforward to isolate a vapor cell from environmental perturbations (for example, placing a cell inside a magnetic shield efficiently suppresses stray magnetic fields). At the same time, the cell itself can be designed so as not to disturb its magnetic or electric environment, a desirable property for non-invasive sensors. Of course, atomic motion introduces several important restrictions to the experimental arrangement -- but we will save those for later discussion as detailed below. 

This paper is organized as follows:  Sec.~\ref{Section:concept} introduces the basic framework for describing EIT and discusses its basic properties. Sec.~\ref{Section:EITinvapor} discusses the effects of atomic motion on EIT. Sec.~\ref{Section:designEIT} discusses important considerations and trade-offs of an atomic vapor-based EIT experiment, while Sec.~\ref{section:buildEIT}  provides practical recommendations for EIT experimental realizations. Finally, Sec.~\ref{Section:conclusions} gives a brief overview of basic EIT applications.

\section{EIT: concept and basic properties}
\label{Section:concept}

\subsection{Dark state}
\noindent There are several ways to explain EIT, but probably the most intuitive one involves the introduction of a so-called ``dark state'' - a quantum superposition of the atomic levels that does not interact with both laser fields. A traditional EIT arrangement includes a three-level atomic (or atom-like) system, in which two of the levels are coupled to the common third level via two near-resonant electromagnetic fields, as shown in Fig.~\ref{fig:EITconfigs}. Assuming that each optical field interacts with only its corresponding transition, the interaction Hamiltonian for such system can be written as
 \begin{equation}
      \hat{H}=
 \begin{pmatrix} -\hbar\omega_{13} & 0 & -\mu_{13} E_1\\ 0 & -\hbar\omega_{23} &  -\mu_{23} E_2 \\ -\mu_{13} E_1 & -\mu_{23} E_2 & 0 \end{pmatrix},\label{3level-ham}
 \end{equation}
 where $\omega_{13}$ and $\omega_{23}$ are the frequency of the corresponding atomic transitions (here we assumed the energy of the state $|3\rangle$ to be zero), $E_{1,2}=\tilde{E}_{1,2}\mathrm{exp}(-i\nu_{1,2}t)+c.c.$ are the electromagnetic fields interacting with each atomic transition, and $\mu_{ij}$ is the dipole moment of the corresponding atomic transition.
 Since we will be mostly interested in the steady-state or slowly varying atomic evolution, we can apply the rotating wave approximation and remove all fast oscillating terms. Under this approximation, and moving to a frame rotating at the fields' frequencies, the Hamiltonian $\hat H$ can be rewritten as
\begin{equation}
      \hat{H}_\mathrm{R}=\begin{pmatrix} -\hbar\Delta_1 & 0 & -\mu_{13} \tilde{E}_1\\ 0 & -\hbar\Delta_2 &  -\mu_{23} \tilde{E}_2 \\ -\mu_{13} \tilde{E}^*_1 & -\mu_{23} \tilde{E}^*_2 & 0 \end{pmatrix}. \label{3level-ham-RWA}
 \end{equation}
Here, $\Delta_i = \nu_i - \omega_{i3}$ is the one-photon detuning of each laser from its corresponding atomic transition.

\begin{figure}
    \centering
    \includegraphics[width=1.0\columnwidth]{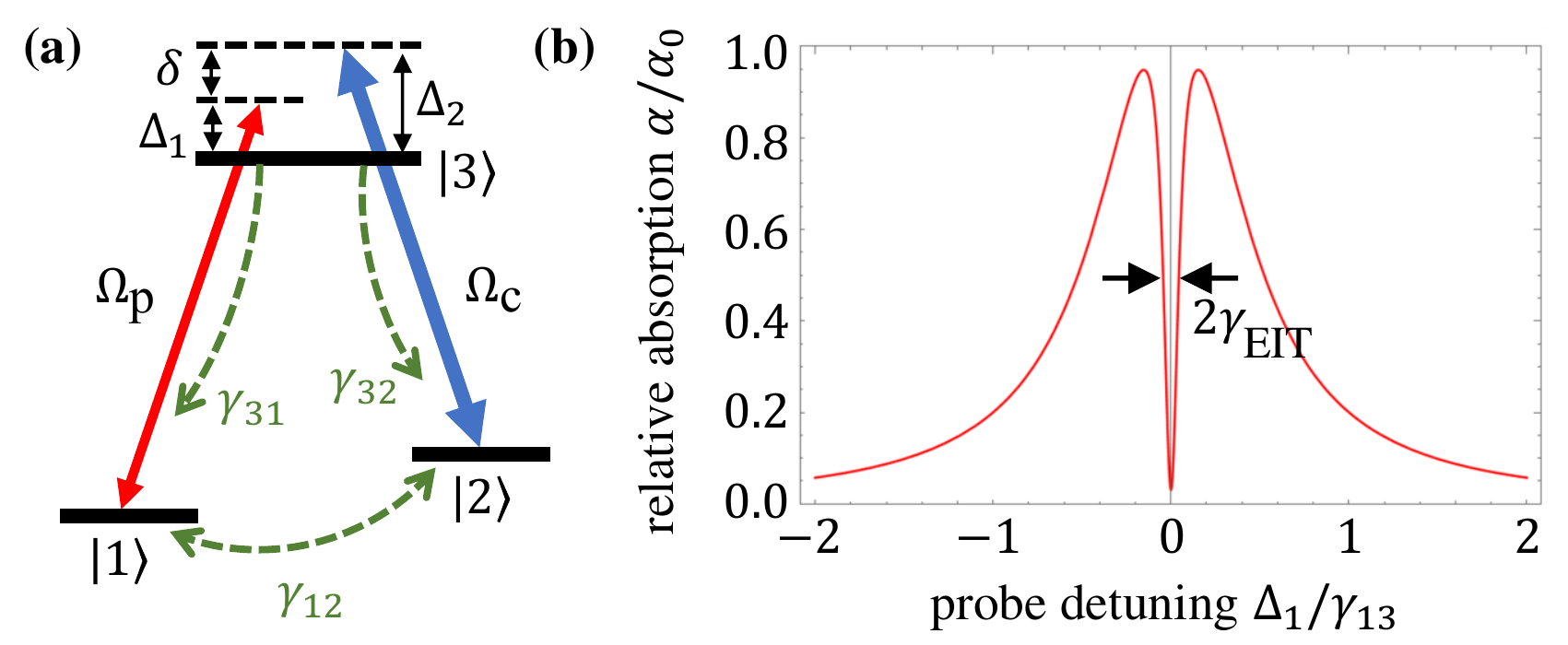}
    \caption{\textbf{Resonant EIT spectrum.} (a) A three-level $\Lambda$ system, considered in the calculations below. (b) Example of the narrow transmission resonance within the probe field homogeneous absorption profile due to EIT for two different values of the control field. In this example $\gamma_{12}=0.001\gamma_{13}$, $ \Delta_2=0$ , and $\Omega_\mathrm{p}=0.001\gamma_{13}$.}
    \label{fig:LambdaEIT}
\end{figure}

A peculiar property of such a Hamiltonian is the existence of a zero-eigenvalue eigenstate, such that $\hat{H}_\mathrm{R} |D\rangle =0$. This state is thus decoupled from the interactions with the optical fields. In the following discussion we focus on a $\Lambda$ interaction scheme, shown in Fig.~\ref{fig:LambdaEIT}, since its dark state involves two long-lived ground states, typically resulting in the most dramatic modification of the optical properties. However, all the procedures are very similar for other schemes. Also, in the majority of experiments one optical field is used as a probe of atomic properties, while the other serves mainly for control purposes; thus, we will use the ``probe'' and ``control'' nomenclature here to describe the two optical fields, and define their Rabi frequencies as $\Omega_\mathrm{p}=\mu_{13} \tilde{E}_1/\hbar$ and $\Omega_\mathrm{c}=\mu_{23} \tilde{E}_2/\hbar$, correspondingly.

  It is easiest to find $|D\rangle$ for the resonant conditions, in which each laser is tuned exactly on resonance $\nu_1=\omega_{13}$ and $\nu_2 = \omega_{23}$, \textit{i.e.} $\Delta_1=\Delta_2=0$, although it exists even for non-zero laser detunings. Notably, $|D\rangle$ is a superposition of only two atomic states $|1\rangle$ and $|2\rangle$:
\begin{equation}
    |D\rangle = \left( \Omega_\mathrm{p}|2\rangle - \Omega_\mathrm{c} |1\rangle\right)/\Omega, \label{dark_state}
\end{equation} 
where  $\Omega=\sqrt{| \Omega_\mathrm{p}^2|+|\Omega_\mathrm{c}^2|}$.  It is also important to note that a dark state exists for any three-level configuration, regardless of exact values of relative energies of the chosen atomic states or the strengths of the optical fields, down to the  single-photon level in a fully quantum EIT treatment~\cite{Kuang_2003}. In case of strong control and weak probe fields ($\Omega_\mathrm{p} \ll \Omega_\mathrm{c}$), most of the atoms will still be found in the state $|1\rangle$, similar to the case of incoherent optical pumping. That is one of the fascinating features of the dark state: if atoms are definitely in the state $|1\rangle$ they strongly absorb the probe field; however, when in a dark state they become ``invisible'', leading (in principle) to a complete transparency, even though they are still \emph{mostly} in the same state $|1\rangle$.

To take full advantage of the EIT effect, the levels are chosen such that the states $|1\rangle$ and $|2\rangle$ have a longer lifetime than the state $|3\rangle$. Especially in the $\Lambda$ configuration, the first two states are chosen among the ground states manifold. In this case, an atom in the dark state cannot be promoted into the electronic excited states, prohibiting the fluorescence and making the atom ``dark'' to an external observer (which is historically the origin of the term ``dark state''~\cite{arimondo'76}). At the same time, the absence of the spontaneous emission removes the dominant optical loss mechanism, so the laser fields can traverse the resonant atomic medium without any absorption -- experiencing the transparency induced by the presence of a control electromagnetic field. 

By analogy to the ``dark state'', we can introduce an orthogonal ``bright'' superposition $ |B\rangle = \left( \Omega_\mathrm{p}^* |1\rangle + \Omega_\mathrm{c}^*|2\rangle \right/\Omega$ that is coupled to the excited state $|3\rangle$, so that the interaction Hamiltonian (for $\Delta_1=\Delta_2=0$) may be written as
\begin{equation}
    \hat{H}_\mathrm{R} = \hbar \Omega \left( |3\rangle \langle B| + |B\rangle \langle 3| \right).
\end{equation}
It is easy to see that the relative phase of the atomic superposition is critical to ensure the non-interaction condition for the dark state. For example, if the phase of one of the fields is suddenly flipped by $\pi$, thus changing the minus sign in the dark state to plus, the atoms temporarily become absorbing, until the atomic coherence is adjusted to the new conditions~\cite{sautenkovJMO2009}. Using this argument, we can qualitatively explain the spectral width of the EIT resonance, although accurate derivation requires the density-matrix formalism described in the next subsection. 

Above, we specifically assumed that the frequencies of each optical field match exactly the frequencies of the corresponding transitions. One can show that even for non-zero laser detunings $\Delta_{1,2}$, the steady-state dark state of Eq.~(\ref{dark_state}) exists for the zero two-photon detuning $\delta=0$, where we define $\delta = \Delta_1-\Delta_2= \nu_1-\nu_2-\omega_{12}$ as the mismatch between the two-photon transition frequency and the frequency difference between states $|1\rangle$ and $|2\rangle$. If a small non-zero two-photon detuning $\delta$ is introduced, the state of atoms initially prepared in $|D\rangle$ evolves in time as
\begin{equation}
    |D_\delta(t)\rangle = \left( \Omega_\mathrm{p} |2\rangle - e^{i\delta\cdot t}\Omega_\mathrm{c} |1\rangle \right)/{\Omega}, \label{dark_state_time}
\end{equation}
causing the sign of the dark-state phase to slowly change. Since in reality atoms cannot maintain their coherence forever, the dark state can exist only for a finite lifetime $\tau_\mathrm{coh}$. So if the two-photon detuning is small, such that the accumulated phase $\delta\cdot\tau_\mathrm{coh}$
is negligible, the dark state stays largely non-interacting, and EIT is preserved. But as detuning increases, the effect of the phase evolution becomes more pronounced. In fact, we can roughly estimate the spectral EIT width to be inversely proportional to the dark-state lifetime by setting  $\delta\cdot\tau_\mathrm{coh}\approx \pi/2$. Such an estimate is quite accurate in the limit of very weak optical fields. We can also use this model to qualitatively explain the so-called power-broadening effect: the increase of the EIT linewidth with the power of the optical fields. Equation (\ref{dark_state_time}) assumes free evolution of the atomic state, which works well for weak optical fields;  the stronger the fields are, the larger is the probability that the evolving state is rephased by the repeated interaction, thus attenuating the free phase evolution and consequently increasing the transparency tolerance to non-zero detuning. 

\subsection{Note on terminology: CPT vs EIT vs Raman}
\noindent One of the difficulties with a literature search on EIT-related research is the different terminology used. For example, the two-photon transmission resonances can be referred to as electromagnetically-induced transparency (EIT), coherent population trapping (CPT), dark resonances, and Raman resonances. Moreover, different people sometimes put slight differentiation between each of these terms, so here we outline what we perceive as the most common definitions. 

CPT is often referred to as the experimental arrangement involving a $\Lambda$-system with two long-lived energy levels (typically two hyperfine or Zeeman ground state sublevels) and two optical fields of comparable strength. In this case atoms are ``trapped'' in a quantum superposition with near-maximum coherence, a process which can be considered as a generalization of the optical pumping process. Such a configuration is most common for metrology applications, such as CPT-based atomic clocks, magnetometers, etc.~\cite{vanier05apb,kitchingAPR2018}.

EIT then is a more general case in which the transmission of a resonant optical field is enhanced by means of another optical field, particularly without any reduction of the atomic population in the initial atomic state. This effect can be realized in any three-level system, and, in principle, for arbitrary values of the optical fields. However, most often EIT experiments imply a strong control optical field $\Omega_\mathrm{c}$ and a weak probe optical field $\Omega_\mathrm{p}$. In this arrangement, EIT looks the most ``counterintuitive'', especially in the ladder system, where adding a strong control field between nominally empty excited states changes the probe absorption dramatically without noticeably changing atomic populations. Indeed, according to the dark-state Eq.~(\ref{dark_state}), for $\Omega_\mathrm{p}\ll \Omega_\mathrm{c}$, the population of the state $|1\rangle$, coupled to the weaker optical field, is $|\Omega_\mathrm{c}^2|/(|\Omega_\mathrm{c}^2|+|\Omega_\mathrm{p}^2|) \approx 1$. This regime is most relevant to quantum information applications, in which EIT is used for realization of the strong coupling between quantum optical fields (particularly, single photons) and long-lived atomic states, \textit{e.g.}, for slow light and quantum memory~\cite{lukin03rmp}. 

EIT is of course a small subset of general two-photon Raman processes. However, in the context of light-atom interaction, Raman resonances often refer to the case of narrow absorption resonances appearing in a $\Lambda$ system when the two optical fields are detuned from the excited state, while maintaining the two-photon resonance (we will consider the effect of the laser detuning on EIT below). In the last decade, such far-detuned Raman resonances became a viable alternative to EIT for quantum memory applications~\cite{Walmsley01052015}.

\subsection{Density matrix description of the EIT}
\noindent While the concept of the dark state provides an intuitive insight into the nature of EIT, the accurate description of this process requires proper account of the decoherence processes for both optical transitions and, more importantly, the atomic coherence associated with the dark state.  Since the wave-function formalism is not adequate for describing quantum systems in the presence of decoherence, we shall utilize the density matrix formalism. 
In this formalism, the evolution of the atomic state matrix $\hat \rho$ under the action of the Hamiltonian $\hat{H}_\mathrm{R}$, is described by the Maxwell-Bloch equation~\cite{scullybook}
\begin{equation}
    \frac{d\hat\rho }{dt} = -\frac{i}{\hbar}\left[\hat{H}_\mathrm{R}, \hat \rho\right] + \mathcal{L}_\Gamma[\hat\rho], 
\end{equation}
where 
the superoperator $\mathcal{L}_\Gamma$ encompasses all the decoherence effects.  We will discuss the specific decoherence effects associated with various aspects of the environment~\cite{XiaoEITwidthreview2009} in detail in Section \ref{Section:EITinvapor}. For now we introduce the general decoherence rates phenomenologically: $\gamma_i$ is the population decay rate of the $i$th state (in case of a state having more than one decay channel, $r_{ij}$ is the branching ratio to the state $j$), and $\gamma_{ij}$ is the decoherence rate of the corresponding off-diagonal matrix element $\rho_{ij}$. Here we will also assume a closed system ({\it{i.e.}}, there is no population exchange outside of the three atomic levels), although it has been shown that the corresponding calculations for the open system result in a very similar outcome~\cite{lee'03}. Finally, since the $\Lambda$ system, shown in Fig.~\ref{fig:LambdaEIT}, is the more common EIT configuration, in the following we will assume that the states $|1\rangle$ and $|2\rangle$ are sublevels of the ground electronic state hence experience no spontaneous emission, and are both coupled to the common excited electronic state $|3\rangle$. In this case, the time evolution equations for the density matrix elements are
\begin{eqnarray}
\dot\rho_{11}&=&r_{31}\gamma_3-i\Omega_\mathrm{p}\rho_{13}+i\Omega_\mathrm{p}^*\rho_{31} \nonumber \\
\dot\rho_{22}&=&r_{32}\gamma_3-i\Omega_\mathrm{c}\rho_{23}+i\Omega_\mathrm{c}^*\rho_{32} \nonumber \\
\rho_{33}&=& 1-\rho_{11}-\rho_{22} \nonumber \\
\dot\rho_{21}&=&-(\gamma_{12}-i\delta)\rho_{21}-i\Omega_\mathrm{p}\rho_{23}+i\Omega_\mathrm{c}^*\rho_{31} \nonumber \\
\dot\rho_{31}&=&-(\gamma_{13}-i\Delta_1)\rho_{31}+i\Omega_\mathrm{c}\rho_{21}+i\Omega_\mathrm{p}(\rho_{11}-\rho_{33}) \nonumber \\
\dot\rho_{32}&=&-(\gamma_{23}-i\Delta_2)\rho_{32}+i\Omega_\mathrm{p}\rho_{12}+i\Omega_\mathrm{c}(\rho_{22}-\rho_{33}).  \label{rho_eqns}
\end{eqnarray} 
These equations can provide the exact solution for any values of experimental parameters, but in general can only be solved numerically. In this section we will consider only the steady-state solution to analyze the main characteristics of the EIT transmission resonances. Even then, though Eqs.~(\ref{rho_eqns}) become a system of linear equations and can be solved analytically, the resulting expressions are rather cumbersome. So here we analyze the most common case of a strong control field and a weak probe field $\Omega_\mathrm{p} \ll \Omega_\mathrm{c}$, in which the system's response to the probe field is linear. 

In the weak-probe regime, it is convenient to use the perturbative approach to the solution, keeping only the linear terms in $\Omega_\mathrm{p}$. To zeroth-order approximation, we can assume that all atomic population is optically pumped into the state $|1\rangle$ (assuming that the control field is sufficiently strong to provide efficient optical pumping): $\rho_{11}^{(0)}=1$, and $\rho_{22}^{(0)} = \rho_{33}^{(0)}=0$. Also $\rho_{23}^{(0)}=0$, since this is the coherence between two empty states. Substituting these values into the right hand side of Eqs.~(\ref{rho_eqns}) and keeping only linear terms in $\Omega_\mathrm{p}$ substantially simplifies the situation since only two equations remain:
\begin{eqnarray}
0&=&-\Gamma_{12}\rho_{21}+i\Omega_\mathrm{c}^*\rho_{31} \nonumber \\
0&=&-\Gamma_{13}\rho_{31}+i\Omega_\mathrm{c}\rho_{21}+i\Omega_\mathrm{p},  \label{rho_eqns_simp}
\end{eqnarray} 
where we use $\Gamma_{12}=\gamma_{12}-i\delta$ and $\Gamma_{13}=\gamma_{13}-i\Delta_1$. This leads to very simple and elegant expressions for the ground state and optical coherences
\begin{eqnarray}
\rho_{21}&=&-\frac{\Omega_\mathrm{p}\Omega_\mathrm{c}^*}{\Gamma_{12}\Gamma_{13}+|\Omega_\mathrm{c}|^2} \\
\rho_{31}&=&i\Omega_\mathrm{p}\frac{\Gamma_{12}}{\Gamma_{12}\Gamma_{13}+|\Omega_\mathrm{c}|^2}  \label{rho_solve}.
\end{eqnarray} 
Then the probe linear susceptibility $\chi_\mathrm{p}$ of an ensemble of $N$ atoms per unit volume is
\begin{equation}
    \chi_\mathrm{p}(\Delta_1,\delta) = \frac{N\mu_{13}^2 }{\hbar\epsilon_0}\frac{\rho_{31}}{\Omega_\mathrm{p}}=i\frac{N\mu_{13}^2}{\hbar\epsilon_0}\frac{\Gamma_{12}}{\Gamma_{12}\Gamma_{13}+|\Omega_\mathrm{c}|^2}.\label{chi_EIT_gen}
\end{equation} 

It is easy to see that, when no ground state coherence exists between the states $|1\rangle$ and $|2\rangle$, $\chi_\mathrm{p}$ reverts back to its value for a two-level system $\chi_\mathrm{p}^{(\textrm{2-level})}(\Delta_1) =i\frac{N\mu_{13}^2}{\hbar\epsilon_0\Gamma_{13}}$. It is convenient to define $\alpha_0$, the unsaturated resonant absorption coefficient (field absorption per unit length) in the absence of EIT, as:
\begin{equation}
    \alpha_0 =k_\mathrm{p}/2\chi_\mathrm{p}^{(\textrm{2-level})}(0)= \frac{k_\mathrm{p} N \mu_{13}^2}{2\hbar\epsilon_0\gamma_{13}},
    \label{eq:alpha0}
\end{equation}
where $k_\mathrm{p}=2\pi\nu_\mathrm{p}/c$ is the probe field's wavevector in vacuum. Then, the susceptibility in Eq.~(\ref{chi_EIT_gen}) can be written as
\begin{equation}
    \chi_\mathrm{p}(\Delta_1,\delta) =i\alpha_0\frac{2\gamma_{13}}{k_\mathrm{p}} \frac{\Gamma_{12}}{\Gamma_{12}\Gamma_{13}+|\Omega_\mathrm{c}|^2}.\label{chi_EIT_gen1}
\end{equation} 

In the ideal case of no ground-state decoherence $\gamma_{12}=0$ and zero two-photon detuning $\delta=0$ ({\it{i.e.}}, for $\Gamma_{12}=0$), the susceptibility completely vanishes, resulting in $100\%$ transparency for the probe field. This result is particularly counter-intuitive for resonant optical fields, as one expects the strongest resonant absorption  due to large atomic population in the state $|1\rangle$. This is the origin of the name for electromagnetically-induced transparency.

\subsection{Near-resonant fields, EIT}
\noindent Let us first consider the case of the probe laser tuned exactly to the atomic resonance $\Delta_1=0$ but allow a small two-photon detuning  $\delta \ll \gamma_{13}$. In this case we can derive the canonical expression for the EIT susceptibility by substituting $\Gamma_{13}=\gamma_{13}$ and $\Gamma_{12}=\gamma_{13}-i\delta$ in Eq.~(\ref{chi_EIT_gen}):
\begin{equation}
    \chi_\mathrm{p}^{\binom{\text{on}}{\text{res.}}}(\delta) = i\alpha_0\frac{2}{k_\mathrm{p}}
    \frac{\left[\gamma_{12}\geit+\delta^2\right]-i\delta|\Omega_\mathrm{c}|^2/\gamma_{13}}{\geit^2+\delta^2}, \label{chi_EIT_res}
\end{equation}
where 
\begin{equation}
   \geit=\gamma_{12}+|\Omega_\mathrm{c}|^2/\gamma_{13}. 
   \label{gammaEIT}
\end{equation}
The absorption coefficient of the probe field in this case is
\begin{equation}
    \alpha_\mathrm{p}(\delta)= \alpha_0\frac{\gamma_{12}\geit+\delta^2}{\geit^2+\delta^2}.\label{abs_EIT}
\end{equation}
 At the exact two-photon resonance $\delta=0$, the probe absorption is suppressed by the factor
 \begin{equation}
    \frac{\aeit}{\alpha_0} = \frac{\gamma_{12}}{\geit}=\frac{\gamma_{12}}{\gamma_{12}+|\Omega_\mathrm{c}|^2/\gamma_{13}}.
 \end{equation} 
 Realistically, we can approach almost complete transparency in the limit of the strong control field $|\Omega_\mathrm{c}| \gg \sqrt{\gamma_{12}\gamma_{13}}$, resulting in the vanishing absorption suppression factor $\frac{\aeit}{\alpha_0} = \frac{\gamma_{12}\gamma_{13}}{|\Omega_\mathrm{c}|^2}\rightarrow 0$.  

Equation (\ref{gammaEIT}) describes the linewidth of the EIT transmission resonance $\geit$. It is easy to see that for a very weak control field, the resonance width is limited by the decoherence rate $\gamma_{12}$, which can in principle be very small, especially in the case of a $\Lambda$ configuration. As the control power increases, the EIT resonance broadens proportionally. For most practical applications, the balance between such power broadening (a narrow resonance needs lower control power) and the absorption suppression factor (higher transmission needs higher control power) determines the optimal control field parameters. Not surprisingly, for many applications the optimal operation conditions correspond to the case when the power broadening term equals the natural decoherence term.

\subsection{Off-resonant fields, EIT and Raman absorption}
\noindent Let us now consider another limiting case in which the probe field is detuned relatively far away from the corresponding atomic transition, such that $\Delta_1 \gg \gamma_{13}$. To analyze the probe absorption for different control-field detunings, it is convenient to rewrite Eq.~(\ref{chi_EIT_gen}) as:
\begin{equation}
    \chi_\mathrm{p}^{\binom{\text{off}}{\text{res.}}}(\Delta_1,\delta) = i\alpha_0\frac{2\gamma_{13}}{k_\mathrm{p}\Gamma_{13}}\left(1-\frac{|\Omega_\mathrm{c}|^2}{\Gamma_{12}\Gamma_{13}+|\Omega_\mathrm{c}|^2}\right). \label{chi_EIT_off}
\end{equation}
Here we can easily identify the first term as a resonant probe interaction, while the second term describes the control field effect. One can check that the largest relative contribution from the second term happens near the two-photon resonance $\delta \ll \Delta_1$. Then, taking into account  $\Delta_1 \gg \gamma_{13}, \gamma_{12}$, we can simplify the expression for the off-resonant probe susceptibility as:
\begin{equation}
    \chi_\mathrm{p}^{\binom{\text{off}}{\text{res.}}}(\Delta_1,\delta) \simeq  i\alpha_0\frac{2}{k_\mathrm{p}}\frac{\gamma_{13}}{\gamma_{13}-i\Delta_1}+i\alpha_0\frac{2}{k_\mathrm{p}}\frac{|\Omega_\mathrm{c}|^2/\Delta_1}{\gamma_\mathrm{R}-i(\delta -\delta_\mathrm{R})}. \label{chi_EIT_off1}
\end{equation}
Here again the first term is the residual linear susceptibility, while the second term corresponds to a two-photon Raman absorption resonance with the width $\gamma_\mathrm{R}=\gamma_{12}+\gamma_{13}|\Omega_\mathrm{c}|^2/\Delta_1^2 \ll \gamma_{13}$, shifted from the exact two-photon resonance by $\delta_\mathrm{R} =  |\Omega_\mathrm{c}|^2/\Delta_1$. 

\begin{figure}
    \centering
    \includegraphics[width=0.85\columnwidth]{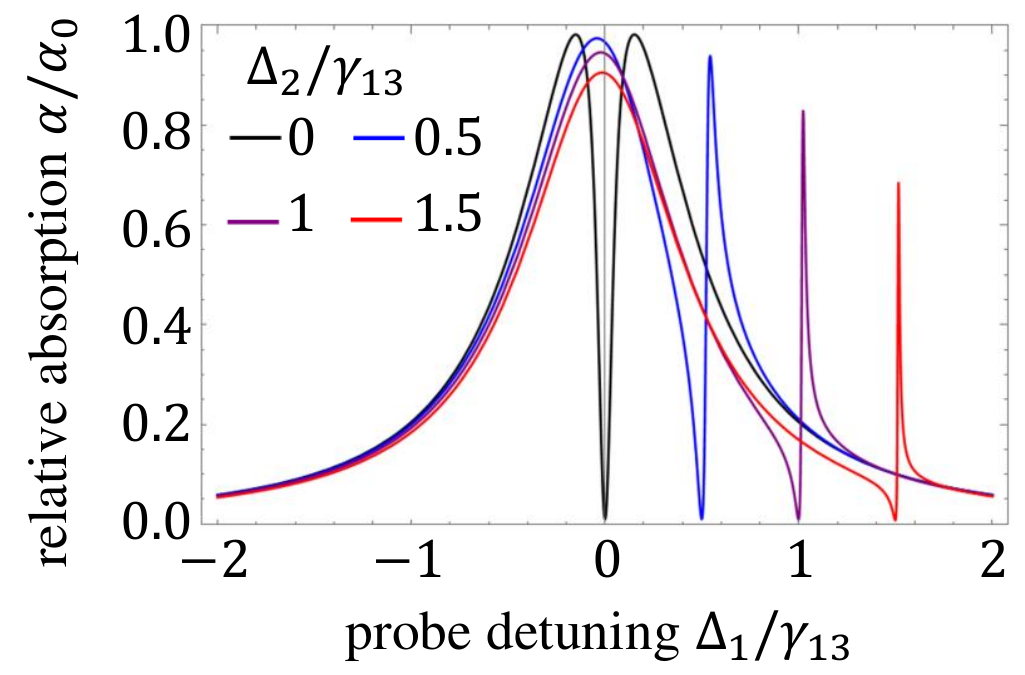}
    \caption{\textbf{Variation of the two-photon resonance lineshape for different control field detunings $\Delta_2$}. In this example $\gamma_{12}=0.001\gamma_{13}$, $\Omega_\mathrm{p}=0.001\gamma_{13}$, and $\Omega_\mathrm{c}=0.3\gamma_{13}$.}\label{Fig:Raman}
\end{figure}

In the general case of the non-zero one-photon detuning, as shown in Fig. \ref{Fig:Raman}, the EIT resonance, which is symmetric when the probe field is tuned exactly to the optical transition, starts to become asymmetric with the laser detuning. Upon further increase of the one-photon detuning beyond the natural linewidth $\gamma_{13}$, the EIT feature transforms into a predominantly absorption resonance (albeit always accompanied by a transparency feature at $\delta=0$ as seen in Fig. \ref{Fig:Raman}). Throughout this transformation, the lineshape of the two-photon resonance can be well-described by a generalized Lorenzian:
\begin{equation}
    \alpha(\delta) = \tilde{\gamma}\frac{A\tilde\gamma+B(\delta-\tilde\delta)}{\tilde\gamma^2+(\delta-\tilde\delta)^2+C}, \label{EIT_genLor}
\end{equation}
where all the parameters $A,B,C,\tilde\gamma,\tilde\delta$ are functions of the one-photon detuning $\Delta_1$. Analytical expressions for all of them straightforwardly follow from Eq.~(\ref{chi_EIT_gen}) and are calculated in \cite{mikhailov04pra,ZanonPRA2011}, but are rather cumbersome. 

\subsection{Ladder scheme and more complex systems}
\noindent In the ladder EIT scheme, the probe field couples the ground state $|1\rangle$ to the excited state $|3\rangle$, and the strong control field is applied between the two excited states $|2\rangle$ and $|3\rangle$, as shown in Fig.~\ref{fig:EITconfigs}(c). The general solution in this case is identical to that of the $\Lambda$ scheme, with the main difference being that now the two-photon coherence $\rho_{12}$ is between the ground and a highly excited state, and thus its decoherence rate is primarily driven by the radiative decay rate of the excited state $\gamma_2$ (and $\Gamma_{12}=\gamma_2/2-i\delta$), where now the two-photon detuning is defined as $\delta =\Delta_1+\Delta_2$. For that reason the ladder EIT is particularly powerful if the highly excited state has a long lifetime, which is the case for Rydberg states in alkali-metal atoms~\cite{Mohapatra2007,kubler_coherent_2010}.

Looking beyond a simple three-level system, one can generalize the expression for the probe linear susceptibility in the case of multiple atomic levels and multiple control fields, forming a chain of coupled states. It is easy to obtain an exact linear response for such a multi-level structure using the following prescription:

\begin{equation}
    \chi_\mathrm{p}=\alpha_0 \frac{\gamma}{\gamma-i\Delta+\sum_j \binom{\text{nested}}{\text{resonance}}_j},
\end{equation}
where
\begin{equation}
    {\small\binom{\text{nested}}{\text{resonance}}_j}=\frac{|\Omega_j|^2}{\gamma_j-i\delta_j+\sum_k  \binom{\text{nested}}{\text{resonance}}_{j,k}},~\text{etc.} \nonumber
\end{equation}
Here each term in the summation $\sum_j$ includes contributions from each control field $\Omega_j$ coupled to the same excited state as the probe field. If there are additional optical fields $\Omega_{j,k}$ linked  through one of the ``primary'' control fields $\Omega_j$, their effect is included as the next ``nested'' level, and so on. 

\begin{figure}[h!]
    \centering
    \includegraphics[width=0.3\textwidth]{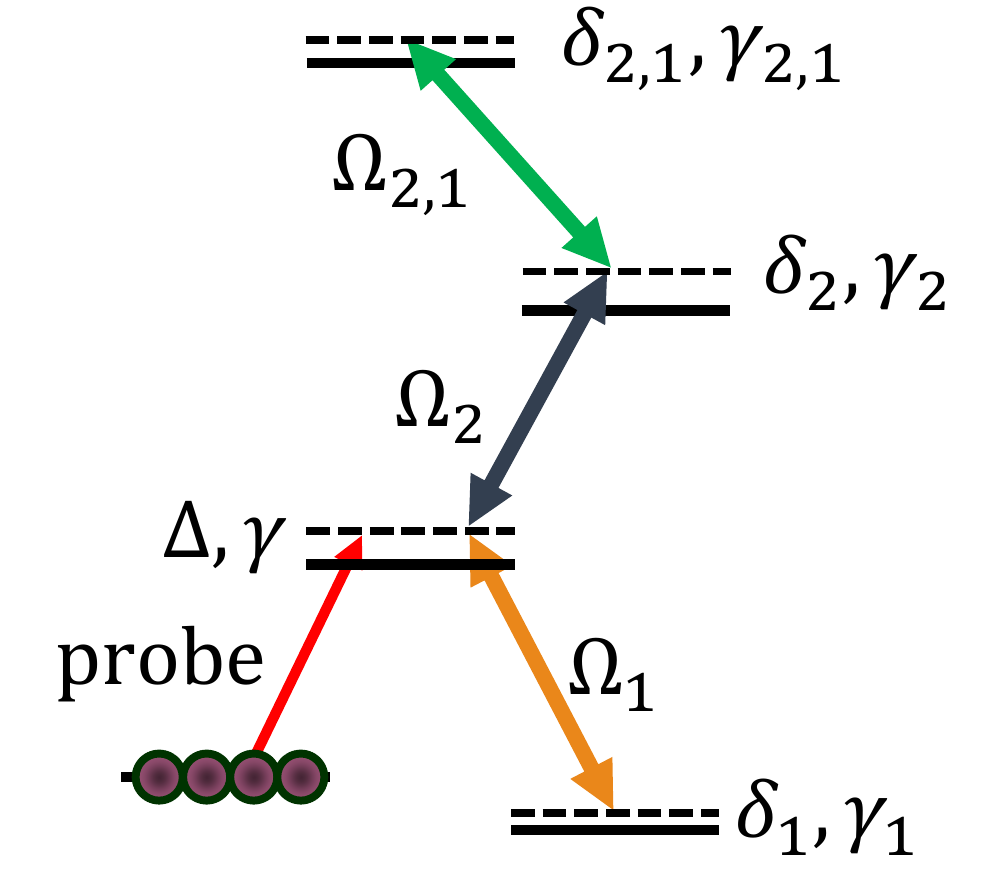}
    \caption{\textbf{Generalized probe susceptibility in a multi-level system.} An example of the relevant parameters for calculating the susceptibility, in the limit of linear response (weak probe), for a combined $\Lambda$ and (4-level) ladder system [see Eq.~(\ref{chi_nested})].}
    \label{fig:multilevel}
\end{figure}
An example of this procedure in action is given in Fig.~\ref{fig:multilevel} for a combined $\Lambda$ and ladder configuration. In this configuration, two ``primary'' control fields $\Omega_1$ and $\Omega_2$ are directly linked with the probe, and the ``secondary'' control field $\Omega_{2,1}$ forms the additional nested resonance. Following the prescription above, one can easily find the probe susceptibility to be
\begin{equation}
    \chi_\mathrm{p}=\alpha_0 \cfrac{\gamma}{\gamma-i\Delta +
    \cfrac{|\Omega_1|^2}{\gamma_1-i\delta_1} + 
    \cfrac{|\Omega_2|^2}{\gamma_2-i\delta_2+ \cfrac{|\Omega_{2,1}|^2}{\gamma_{2,1}-i\delta_{2,1}}}} \label{chi_nested}
\end{equation}

\subsection{Physical interpretation of EIT and Raman resonances based on the dressed-state picture}
\noindent While the counter-intuitive behavior of the two-photon resonances (the lack of absorption when it is expected or, vice versa, strong narrow absorption resonance far off resonance) is part of their charm, we can gain some intuition about their nature by considering an atom dressed by the strong control field. Such a system can be represented by two dressed states with energies $E_\pm$  and corresponding eigenstates $|\pm\rangle$ that are combinations of atomic states $|2\rangle$ and $|3\rangle$. It is particularly interesting to consider two special cases: when the control field is  resonant ($\Delta_2 = 0$) or far-detuned ($\Delta_2 \gg |\Omega_\mathrm{c}|$). 

In the first case (EIT conditions), the dressed states are symmetrically shifted up and down by the same amount from exact atomic resonance:  
\begin{eqnarray}
   E_\pm  & = & \pm \hbar \Omega_\mathrm{c} \nonumber\\
    |\pm\rangle &= &( |2\rangle\pm|3\rangle)/\sqrt{2},
\end{eqnarray}
where we assume $\Omega_\mathrm{c}$ to be real for simplicity. This is a well-known Autler-Townes doublet that can be observed in the absence of phase coherence between control and signal optical fields. In fact, distinguishing between EIT and Autler-Townes mechanisms for some experimental arrangement is a non-trivial task~\cite{AnisimovPRL2011,2013laurat}, since in both cases the presence of a  strong control field results in the reduction of the resonant probe absorption. There is an important distinction, however: in case of EIT the probe light interacts with two otherwise identical states of opposite parity that results in destructive interference of atomic polarizability leading to complete suppression of the probe resonant absorption (at least in the ideal case of negligible spin decoherence) . If the Autler-Townes effect prevails, the reduction of absorption is solely due to the shifts of the absorption lines, so there is always some residual absorption at the resonance, caused by the wings of two overlapping Autler-Townes absorption peaks. 

The dressed state picture is also helpful to understand the presence of the narrow two-photon absorption resonance for a far-detuned $\Lambda$ system. For $\Delta_2 \gg \Omega_\mathrm{c}$, the two dressed states are not at all symmetric:
\begin{eqnarray}
   E_-  & = & - \hbar \Omega_\mathrm{c} - \hbar \Omega_\mathrm{c}^2/\Delta_2, \nonumber\\
    |-\rangle & \simeq & |3\rangle + \Omega_\mathrm{c}/\Delta_2 |2\rangle, \\
    E_+  & = & \hbar \Omega_\mathrm{c}^2/\Delta_2, \nonumber\\
    |+\rangle & \simeq & |2\rangle - \Omega_\mathrm{c}/\Delta_2 |3\rangle. \label{dressed_EIT_off} 
\end{eqnarray}
The negative dressed state $|-\rangle$ consists of mainly the excited atomic state and is responsible for the regular absorption resonance when the probe field is tuned to $|1\rangle \rightarrow |3\rangle$ transition. At the same time, the probe interacting with the positive state $|+\rangle$ is primarily coupled to the long-lived $|2\rangle$ ground state that governs the narrow spectral linewidth of the corresponding absorption resonances, as can be observed in Fig.~\ref{Fig:Raman}. Physically, this transition enables ``tunneling'' of atomic population between the two ground states with minimum involvement of the excited state, causing the optical properties of the probe field susceptibility near this transition to be dominated by the ground state coherence.
Of course, one can use the dressed state formalism to calculate the probe field optical response for any control detuning, although it becomes harder to find an intuitive description of the resulting Fano resonance lineshapes, given by Eq.~(\ref{EIT_genLor}).


\subsection{EIT dynamics: slow and stored light}
\noindent Up to now we have paid particular attention to the atomic absorption resonances. However, the corresponding spectral features in the refractive index are equally exciting. In particular, a power-dependent steep dispersion associated with narrow EIT resonances enables realization of controllable group velocity in a wide range from ``slow'' to ``fast'' light. 

One can calculate the refractive index for the probe field using the real part of the susceptibility calculated in Eq.~(\ref{chi_EIT_gen}): $n=1+\mathrm{Re}(\chi_\mathrm{p})$, although the general expression is rather cumbersome. To find the group velocity under the EIT conditions, however, we can neglect the one-photon detuning $\Delta_1 = 0$ and find the refractive index as a function of $\delta$ at the EIT peak
\begin{equation}
    n(\delta)=1+  \alpha_0\frac{2\gamma_{13}}{k_\mathrm{p}}\frac{\delta} {\geit^2+\delta^2}\frac{|\Omega_\mathrm{c}|^2-\gamma_{12}^2-\delta^2}{\gamma_{13}^2}.
    \label{EITrefindex}
\end{equation}
It is easy to see that exactly on resonance the refractive index is equal to one, but then varies rapidly with the two-photon detuning. In fact, it is possible to obtain the enhanced refractive index by tuning the lasers a little off the exact EIT resonance, while still taking advantage of the reduced absorption~\cite{Zibrov96prl,yavusPRL05}.

The group velocity is determined by the slope of the dispersion curve, which is maximum at the EIT peak $\delta=0$:
\begin{equation}
    v_\text{g}=\frac{c}{n_\text{g}}=\frac{c}{1+\nu_\mathrm{p}\frac{dn}{d\nu_\mathrm{p}}\big|_{\delta=0}}.
    \label{vgroup_def}
\end{equation}
Assuming $\delta\ll\Omega_\mathrm{c},\gamma_{13}$, we can find that the probe group velocity under the EIT conditions is determined by the strength of the control field,
\begin{equation}
    v_\text{g}=\frac{c}{1+ \frac{c\alpha_0}{\pi\gamma_{13}}\frac{|\Omega_\mathrm{c}|^2}{\geit^2}}.
    \label{vgroup_EIT}
\end{equation}
It is clear from Eqs.~(\ref{gammaEIT}) and (\ref{vgroup_EIT}) that by using a weaker control field, the group velocity can be reduced  by many orders of magnitude compared to $c$. This regime is often referred to as ``slow light'' and was demonstrated in 1999 in both cold and hot atoms~\cite{hau99,kash99,budker99}. The demonstrations of optical pulses propagating with speeds of a few tens of meters per second in both ultracold and hot atoms under EIT conditions attracted a lot of attention among scientists as well as the general population. Since then, this effect has been demonstrated in a wide variety of systems and has been considered for many applications. More detailed descriptions and experimental realizations of EIT-based slow light experiments in Rb vapor are discussed in Refs.~\cite{milonni_book,novikovaLPR12,derose2022producing}.

Similarly, tuning to the bottom of the Raman absorption resonance can provide equally large but negative dispersion $dn/d\nu_\mathrm{p}$, making the group velocity exceed the speed of light or even reach negative values (the ``fast'' or ``superluminal'' regime)~\cite{mikhailov04pra_grp_vel}. The initial reports of the fast light observations~\cite{wang2000nature}, in which a pulse after interacting with atoms seemed to emerge ahead of its copy propagating in vacuum, initiated a lot of discussion regarding possible causality violation, as well as various definitions of speed with which information can be transferred in dispersive materials. Since then, the superluminal regime has been observed using both absorptive and gain resonances and has been proposed for improving the sensitivity of optical gyroscopes and white-light cavities~\cite{shahriar2007pra_fast_gyro}.

Probably the most consequential EIT-related application is its role in development of quantum information technologies, and in particular quantum memories. Since two-photon interaction allows strong coupling between a weak optical field and a long-lived atomic coherence, mediated by the strength of a strong control field, the dynamic variation of this field allows reversible mapping of the quantum state of the probe field onto the collective quantum state of the atomic ensemble, and vice versa~\cite{hau01nature,phillips01prl,zibrov02prl}. 

The detailed discussion of quantum memory operation is beyond the scope of this manuscript~\cite{lukin03rmp} and is the subject of several in-depth reviews~\cite{euromemory,quantmemoryreviewJMO2016,Ma_2017}. To describe the basic principle of operation for the EIT quantum memory, it is convenient to think of the propagating probe optical field and the atomic ensemble as a single quasi-particle (often referred to as a ``dark state polariton'', or DSP) consisting of coupled photonic and atomic components~\cite{fleishhauerLukinPRL00}. The ratio between the two is determined by the group velocity and, correspondingly, by the strength of the control optical field $\Omega_\mathrm{c}$. Weaker $\Omega_\mathrm{c}$ reduces the group velocity, as predicted in Eq.~(\ref{vgroup_EIT}), and increases the ``weight'' of the atomic component of the dark-state polariton. A sudden reduction of the control field to zero arrests the DSP motion, forcing it to convert completely into atomic coherence. The quantum information is thus stored in the atomic memory for times shorter than the polariton lifetime, which can reach up to seconds for the case of atomic ground-state spin states. This information can then be read-out on-demand by simply restoring the control field, reactivating the DSP motion. Many experiments have demonstrated the effectiveness of such quantum memory for storing quantum optical information carried by probe photon number, polarization, optical angular momentum, and more~\cite{euromemory,quantmemoryreviewJMO2016,Ma_2017}.

Another important aspect of atoms' ability to preserve quantum information is that it enables light-atom entanglement that is proven to be a crucial step in the development of a quantum repeater -- a critical element for long-distance quantum communication. The first proposal for a practical quantum repeater~\cite{DLCZ} used spontaneous off-resonant Raman scattering to simultaneously produce a heralding photon and a correlated collective atomic excitation that can be later converted on-demand into a second photon via EIT-based control field read-out. This process thus produces a pair of entangled photons with controllable time interval that can be used to establish and extend the entanglement between neighbouring nodes of a quantum network with beneficial resource scaling for longer distances. At the same time, this mechanism can also be used to realize an on-demand single photon source.


\section{EIT in atomic vapor}
\label{Section:EITinvapor}
\noindent Having presented the basics of EIT for stationary atoms, we now discuss the effects of thermal atomic motion.
This motion, due to the atoms' high kinetic energy, is arguably the most prominent characteristic of thermal vapor. It results in a time-dependent interaction with the light fields, which is manifested as spectral broadening in the frequency domain or as decoherence in the time domain. To evaluate the importance of different motional effects, one should consider the time of flight $T$ of an atom through the different length scales $l$ in the problem. For example, $l$ might be the wavelength of the excitation fields or the size of a finite excitation volume. The resulting spectral broadening is then given by
\begin{equation}
    \sigma \simeq 1/T.\label{time_of_flight}
\end{equation}
Generally, $T\propto l$ for ballistic motion, and $T\propto l^2$ for diffusive motion (as in a random walk).

\begin{figure*}[t]    \centering
    \includegraphics[width=16.0 cm]{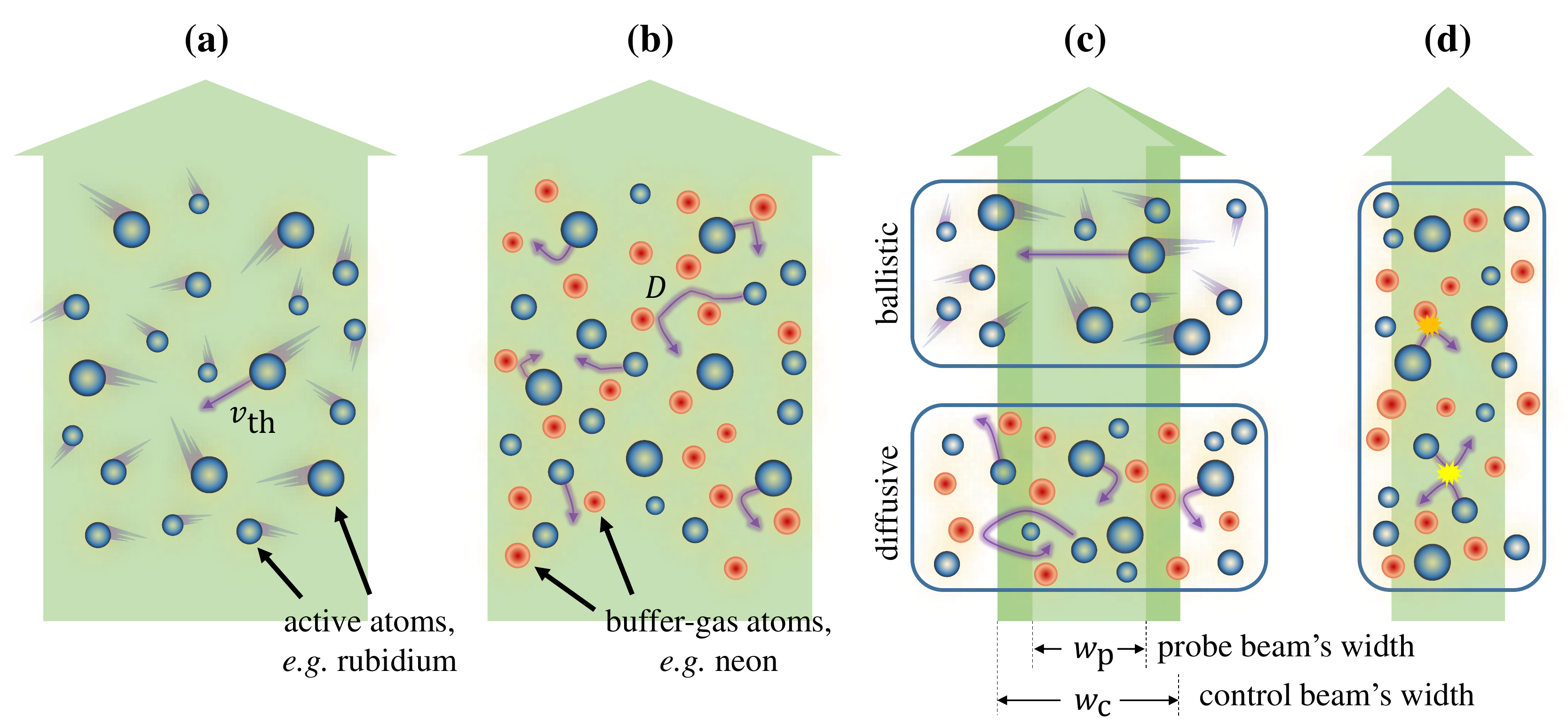}\caption{ \textbf{Motional broadening of EIT in atomic vapor.} (a) Atoms in ballistic motion interacting with an ideal infinite beam (a plane wave). (b) Atoms in diffusive motion in the presence of a buffer gas, interacting with an infinite beam. (c) For finite probe and control beams with waist radii $w_\mathrm{p}$ and $w_\mathrm{c}$, atoms move in and out of the interaction region. (d) Atoms moving in a dense gas experience rapid velocity-changing collisions, potentially also involving (internal) state changes.}
    \label{fig:thermal}
\end{figure*}

Atomic motion affects the EIT lineshape via both `one-photon' and `two-photon' contributions~\cite{FirstenbergPRA2008}. The one-photon contribution refers to spectral broadening and decoherence of the $|1\rangle-|3\rangle$ transition, associated with the optical coherence $\rho_{13}$. This broadening is always present in vapor media, and, while it does not affect the EIT linewidth in principle, it acts to attenuate the absorption cross-section of the media. This attenuation results in a reduced optical depth. 
The two-photon contribution pertains to the overall transition from state $|1\rangle$ to state $|2\rangle$, associated with the (spin) coherence $\rho_{12}$. Motional broadening and decoherence of this transition results in broadening of the EIT linewidth and in reduced contrast of the EIT susceptibility (both real and imaginary parts). In the following, we discuss the different regimes of atomic motion in thermal vapors, as illustrated in Fig.~\ref{fig:thermal}, and their consequences.

\subsection{Motional effects for ballistically moving atoms in an infinite beam} 
\noindent We begin by considering very wide beams, ideally plane-wave fields, as illustrated in Fig.~\ref{fig:thermal}(a). For thermal atoms moving ballistically in such a field, the optical wavelength $\lambda$ sets the only relevant length-scale for motional effects, and the corresponding time-scale is $T\simeq\lambda/\vt$, where $\vt$ is the mean thermal velocity [the root-mean-square (rms) of the one-dimensional velocity distribution]. The general formula (\ref{time_of_flight}) then yields an estimate of $\sigma\simeq\vt/\lambda$ for the expected spectral broadening.
This broadening, known as Doppler broadening, can be alternatively estimated from the Doppler shifts $\delta\rightarrow\delta-\vec{k}\cdot\vec{v}$ each atom experiences in its own reference frame. 
Here $\vec{v}$ is the atom's velocity, $\vec{k}$ is the wavevector ($|\vec{k}|=2\pi/\lambda$), and $\delta$ is the detuning between the laser frequency and the atomic transition frequency.
The Doppler shift results in an inhomogeneous spectral broadening when averaging over the different velocities in the ensemble or over the different wavevectors composing the field.

More precisely, we consider an ensemble of atoms moving in a dilute medium, such that velocity-changing collisions are scarce and the atomic velocities can be considered fixed. The atoms traverse a uniform field with a periodically-modulated phase along the field's propagation direction.
Using the transit-time approach [Eq.~\eqref{time_of_flight}], the Doppler broadening can be evaluated by considering the time it takes an atom to cross $l=\lambda/(2\pi)$, \textit{i.e.} one radian of the phase of the excitation field:
\begin{equation}
\sigma_\text{Dop}=\frac{ \vt}{\lambda/2\pi}=\vt|\vec{k}|.
\label{eq:sig_doppler}
\end{equation}
Typical thermal velocities are in the range $\vt=100-300$ m/s at ambient  temperature, and the optical wavelength is on the range of $0.3-1.5 \mu m$. Therefore, a typical width $\sigma_\text{Dop}$ of the one-photon transition is a few hundreds of MHz. 
Alternatively, using the Doppler-shifts approach, the exact spectral broadening 
is obtained by summing over the atomic velocities, usually following the one-dimensional Maxwell-Boltzmann distribution 
\begin{equation}
w(v)=\frac{1}{\sqrt{2\pi}\vt}e^{-v^2/ 2 \vt^2}.
\end{equation}
Assuming without loss of generality $\vec{k} || \hat{z}$, the averaged susceptibility is given by 
\begin{equation}
\label{eqn:VoigtAbsorption}
    \chi_{\mathrm{ensemble}}(\delta) = \int_{v_z=-\infty}^{\infty} \chi_\mathrm{p}(\delta-kv_z)w(v_z) dv_z.
\end{equation}
Therefore, the lineshape of the Doppler-broadened ensemble is a convolution of Lorentzian and Gaussian profiles, a spectrum known as a Voigt profile. A useful approximation to this profile is the pseudo-Voigt function, which is simply a linear combination of the Gaussian and the Lorentzian functions.

\begin{figure*}[t]    \centering
    \includegraphics[width=16cm]{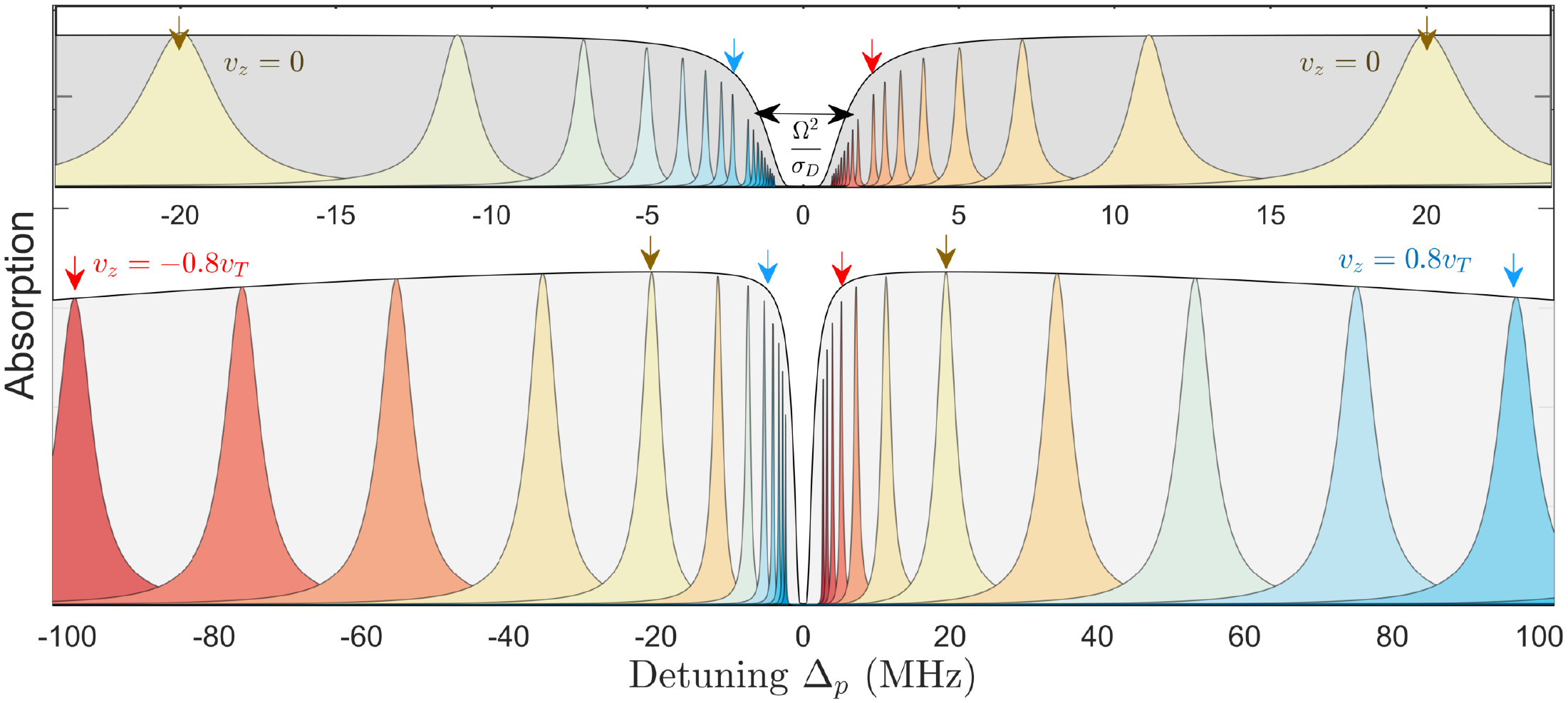}\caption{ \textbf{EIT in thermal vapor under the effect of a significant Doppler broadening.} Bottom panel: a Doppler broadened absorption line is composed of different atomic velocity groups. A narrow EIT line forms at its center,where no velocity group absorbs light. Top panel: a closer view around the EIT window, highlighting the stationary atoms ($v_z=0$) absorption which forms an Autler-Townes doublet, and the EIT linewidth. Here the control Rabi frequency is $\Omega=20$ MHz and the one-photon Doppler width $\sigma_D=230$  MHz.
    \label{fig:v_groups}}
\end{figure*}

We now turn our focus to the EIT resonance. As EIT is a coherent two-photon process, it is dominated by the effective two-photon field $\Omega^*_\mathrm{c}\Omega_\mathrm{p}$. Atomic motion through this field, which has an effective two-photon wavevector $\keff=\kp\pm\kc$, governs the motional broadening of the EIT resonance. Here $(-)$ corresponds to a $\Lambda$ level system and $(+)$ for a ladder system, {\it{i.e.}} depending on whether a photon in absorbed from or emitted into the control field during the excitation of the EIT spin wave. When the fields are similar in wavelength, one can choose the propagation direction of the control and probe (co-propagating or counter-propagating) such that $\keff$ is small. In a $\Lambda$ scheme there could be complete degeneracy (direction and frequency), such that $\keff=0$. Often this degeneracy is lifted by assigning a slight angle of the control field relative to the probe field or by choosing two ground states with different energies due to applied magnetic field or hyperfine splitting. The small yet finite effective excitation wavelength is typically on the order of few $\text{cm}^{-1}$ up to $\text{m}^{-1}$. 

Figure \ref{fig:v_groups} shows how EIT is maintained in an ensemble of atoms in thermal motion. Atoms at different velocities experience large one-photon Doppler shifts leading to an absorption linewidth of a few hundreds of MHz (grey shaded area). The absorption spectrum for each velocity group in the atomic ensemble features a doublet feature akin to Autler-Townes splitting. However, there is a narrow frequency region where none of the atomic velocity groups absorbs light. It is this region where the EIT line is formed, and thus all of the atomic velocity groups take part in the transparency formed. The power-broadened EIT linewidth is thus set by the inhomogeneous broadening of the excited state
\begin{equation}
    \geit=\frac{\Omega_\mathrm{c}^2}{\sigma_\text{Dop}}+\gamma_{12},
    \label{eq:gammaEITsigma}
\end{equation}
 where $\sigma_\text{Dop}$ is the one-photon Doppler width. 
 
 In a Ladder scheme, there is typically a big wavelength mismatch, and $\keff$ cannot be made much smaller than the inverse of that residual wavelength, typically on the order of microns to hundreds of microns. The Doppler shift of the EIT line for a given atomic velocity $v$ is $\delta\rightarrow\delta-\keff\cdot\vec{v}$, which one can plug into the three-level susceptibility expression given by Eq.~\eqref{chi_EIT_res}.
The resulting broadening is known as Residual or two-photon Doppler broadening and is on the order of few MHz to tens of MHZ for ladder-type systems. Finally, we note that such motional dephasing is a special case of the general problem known as inhomogenenous dephasing. As such, it can in principle be reversed and eliminated  \cite{Finkelstein2021}.

\subsection{Diffusive infinite beam}~\label{subsec_diffusion}
\noindent As we have seen, the ballistic motion of the atoms typically leads to substantial broadening and fast decoherence. A prominent solution to this problem in hot vapor is to add a relatively dense, inert buffer-gas. Frequent collisions between the (optically) active atoms and the buffer-gas atoms effectively suppress the ballistic motion, as illustrated in Fig. \ref{fig:thermal}(b). Between collisions, the atomic velocity of the active atoms is constant and still governed by the bare thermal distribution. However, the velocity gets redistributed over several collisions (or even upon a single collision in the so-called `hard collision' limit). The overall effect of these frequent velocity-changing collisions is a diffusive motion of the active atoms. For now, we disregard the effect of the collision on the internal state of the atoms, predominantly causing dephasing of the optical transitions (so-called pressure broadening), which we discuss in Sec.~\ref{subsec_collisional} below.

Given a diffusion coefficient $D$, the transit time through a typical length scale $l$ is given by $l^2/D$, resulting in a spectral broadening of 
\begin{equation}
\sigma=\frac{D}{l^2}.
\label{eq:sig_diffusion}
\end{equation}
This should be compared to the broadening due to ballistic motion $\sigma=\vt/l$, and the actual broadening would be the \emph{smaller} of the two. When $D/l^2<\vt/l$, the diffusion broadening prevails, narrowing the Doppler broadening by the factor $D/(\vt l)$. This effect is known as Dicke narrowing \cite{Dicke1953,FirstenbergPRA2007}. Note that Dicke narrowing occurs when $l$ (usually the relevant wavelength) is larger than the effective mean free-path between collisions $D/\vt$. 

For a single optical transition, the relevant length scale $l=\lambda/(2\pi)$  (on order 100 nm) is typically shorter than the collision mean free-path (a few  microns for 10 Torr of buffer gas), and Dicke narrowing is hard to reach (although it has been demonstrated in a nanoscale vapor cell~\cite{Sargsyan_2016}). For most experimental conditions, the motional contribution to the optical line is therefore usually a Gaussian due to Doppler broadening. 

On the other hand, the EIT transition has a (two-photon) wavelength $2\pi/|\kp\pm\kc|$ that, in the case of the $\Lambda$ level configuration, often surpasses the mean free-path, and hence Dicke narrowing prevails. The EIT line broadening then takes the form of a Lorentzian with a HWHM of $D k^2=D (\kp-\kc)^2$ \cite{shukerPRA2008}. For example, in the case of  so-called hyperfine EIT, shown in Fig.~\ref{fig:EITconfigs}(a), where the frequency difference $c(\kp-\kc)$ is on the order of a few GHz, the two-photon wavelength is a few centimeters. The resulting, Dicke-narrowed, linewidth is a few Hz (for $D\approx 10~\text{cm}^2/\text{s}$ with 10 Torr buffer gas; see Table V in Ref.~\cite{happer72} for the diffusion coefficients in common buffer-gasses). This width is much smaller than that expected without a buffer gas ($\sim 10$ kHz) and usually negligible compared to other broadening mechanisms. 

\subsection{Finite beam effects}
\noindent For a beam of finite width, with a transverse dimension smaller than the extent of the atomic medium, as illustrated in Fig.~\ref{fig:thermal}(c), transverse motion of atoms through the beam results in what is known as transit-time broadening. This spectral broadening can be evaluated according to Eq.~\eqref{time_of_flight} as the inverse of the time of flight through the beam waist
\begin{equation}
  \Gamma_\text{tt}=\frac{\vt}{w_0}~~~\mathrm{or}~~~ \frac{D}{w_0^2},
\end{equation}
where, again, the smaller one prevails.

For a ballistic transverse motion, an exact solution can be obtained by summing susceptibilities for different velocities, similarly to the Doppler broadening due to longitudinal motion \eqref{eqn:VoigtAbsorption}. However, for transverse motion one must also sum over the different wavevectors of the field interacting with the atoms
\begin{equation}
    E_\text{out}(\delta) = \int d^2 k_\perp  E_\text{in}(\delta,\vec{k}_{\perp})e^{ik_zL \int d^2v\, w(\vec{v})\chi_\mathrm{p}(\delta,k_\perp,v)},
\end{equation}
where $E(\delta,\vec{k}_{\perp})$ is the two-dimensional Fourier transform of the transverse envelope of the incoming field. For the general case of EIT, when both the probe and control fields are confined, such a full solution is not trivial. However, it is instructive to consider a more simple case where only the probe field is a finite Gaussian beam. In this case, the solution has a 'cusp'-like lineshape
\begin{equation}
\chi_{\mathrm{ensemble}}(\delta) \propto e^{-|\delta|\frac{w_0}{\vt}},
\end{equation}
with a full-width at 1/e of 
\begin{equation}
    2\Gamma_\text{tt}=2\frac{\vt}{w_0}
\end{equation}
or FWHM of $2(\vt/w_0)\log 2$. The cusp shape can be intuitively explained by the nature of the transit time broadening, where slow atoms, which contribute a narrow-line spectrum, also have a larger weight in the overall spectrum due to the longer period of interaction  with the probe field. For a vast range of experiments with mm-sized beams, this broadening is on the order of tens of KHz, while for experiments with beams focused to a few microns this already yields a spectral broadening of few MHz, on the order of the natural width of the excited state. 

\begin{figure}[bt]    \centering
    \includegraphics[width=0.9\columnwidth]{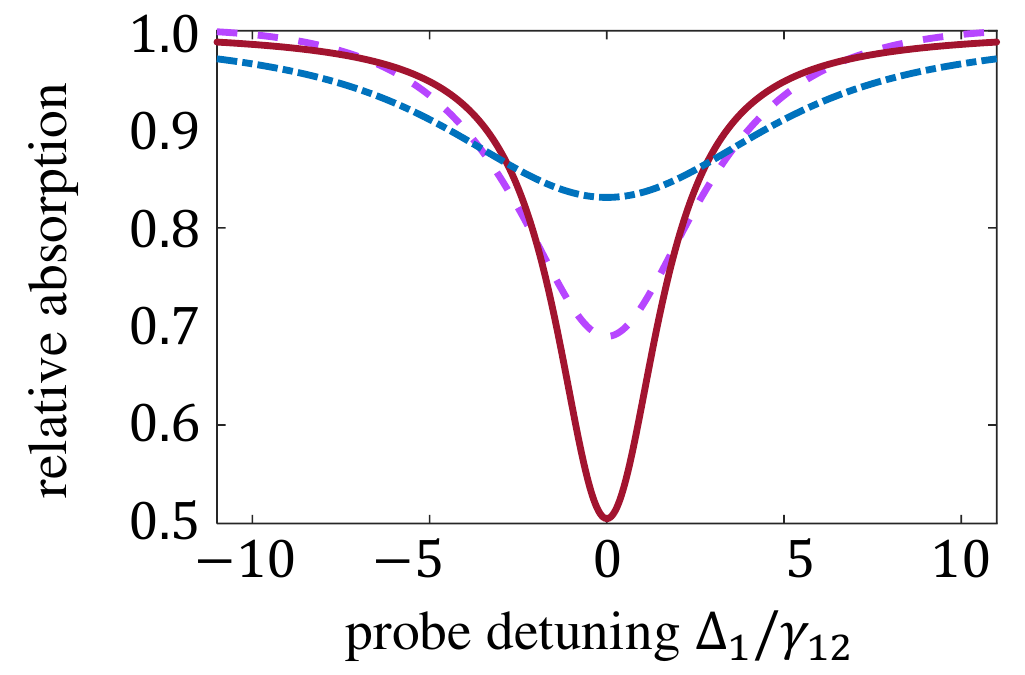}\caption{ \textbf{EIT lineshape for various regimes of motional broadening.}  calculated EIT lineshape with no wavelength mismatch (solid red line), with small wavelength mismatch of $10^{-5}$ of the optical wavelength (dashed blue line) resulting in a residual Doppler broadening, and with a finite beam width (dashed purple line) resulting in transit-time broadening.  In all cases $\gamma_{13} = 3 $ MHz, $\gamma_{12}= 1 $ KHz, and $\Omega_\mathrm{c}=0.5$ MHz.
    \label{fig:lineshapes}}
\end{figure}

In Fig. \ref{fig:lineshapes}, we plot the EIT lineshape for a $\Lambda$ system under different motional regimes and for realistic finite decoherence rate of the ground states $\gamma_{12}$. For a slight angle between the probe and control fields or a slight energy mismatch, such as in the case of hyperfine EIT, the residual Doppler broadening for atoms in ballistic motion results in a Voigt lineshape with reduced contrast (blue dashed line). However, for the same configuration, motion in the diffusive regime can result in a Dicke narrowed line with a width approaching the natural linewidth (solid line). Transit-time broadening due to finite transverse dimensions of the beam results in a cusp lineshape convoluted with the natural Lorentzian lineshape (dashed purple line).

Turning now to discuss the diffusive regime in the transit-time problem, we again differentiate between infinite and finite control fields. In the former case, when only the probe beam is finite, it is enough to (linearly) average over the different wavevector composing the probe field,
\begin{equation}
    E_\text{out}(\delta) = \int d^2 k_\perp  E_\text{in}(\delta,\vec{k}_{\text{p}\perp})e^{ik_{\text{p},z}L \chi_\mathrm{p}(\delta,\vec{k}_{\text{p}\perp})},
\end{equation}
where $\chi_\mathrm{p}(\delta,\vec{k}_{\text{p}\perp})$ includes the Dicke-broadened two-photon linewidth $\geit+ D (k_{\text{p},z} \hat{z}+\vec{k}_{\text{p}\perp}-\kc)^2$. For nearly-degenerate, co-propagating probe and control $k_{\text{p},z} \hat{z}\approx \kc$, the EIT line is overall broadened by $D/w_\text{p}^2$, where $w_\text{p}$ is the probe's waist. 

As before, the calculation becomes more involved if the control beam is finite as well. For a weak control field, a direct convolution over the control's wavevectors can be done. However for strong control fields, when the power broadening $\Omega_\mathrm{c}^2/\sigma_\text{Dop}$ is much larger than $\gamma_{12}$, the (nonlinear) contribution of the control complicates the calculations. For some arrangements, though, one can perform the calculation in the spatial (rather than the spectral) regime. One notable example is the arrangement of identical probe and control fields ($w_\text{p}=w_\text{c}$), with atoms constantly diffusing in and out of the illuminated area. In this regime, the illuminated periods, interrupted by dark periods, can be thought of as a stochastic train of Ramsey spectroscopy sequences. The resulting EIT line has a characteristic sharp peak (cusp or cusp-like shape) with a typical width $\gamma_{12}$, {\it{i.e.}}, it is unaffected by power broadening \cite{XiaoPRL2006,FirstenbergPRA2008}. This effect, known as Ramsey-narrowing, can be explained by the atoms spending a long time outside the beams, in the dark, before the repeated interaction with the optical fields. 

Beyond broadening and narrowing, there are several interesting spatial effects occurring in the finite-probe regime. Traversing the medium as a slow-light polariton, the probe field is directly affected by the motion of the atoms. Particularly on EIT resonance, the atoms effectively carry the field with them. As a result, a uniform drift velocity of the atoms results in transverse drag of the probe field \cite{solomons2020}, while a diffusive motion of the atoms results in spatial diffusion of the field \cite{shukerPRL2008,firstenbergRMP2013,chrikiOptica2019}. Notably, the diffusion here is of a complex quantity (a `coherent diffusion' of both the argument and the phase of the field) and it therefore demonstrates interference phenomena, such as self-similar expansion and contraction of the probe field \cite{pugatchPRL2007,FirstenbergPRL2010,smartsevOE2020}. More intriguing results occur slightly off the EIT resonance, such as negative drag and negative diffraction \cite{firstenbergNP2009,BanerjeeArxiv2021}.

Finally, some atoms reach the enclosing walls of the cell. If the walls are not coated by a spin-preserving material (see Sec.~V below), the atoms eventually bounce from them completely decohered. They therefore cause a transit-time broadening, which can be described as outlined above for a finite beam, with the cell width replacing $w_0$. If the walls are coated, they can manifest, in the ideal case, as a periodic boundary condition. In this case, for degenerate and collinear probe and control fields ($\kp\pm\kc=0$) and assuming the beams cover the entire (finite) cell, the situation is akin to an infinite-beam arrangement, and we expect no motional decoherence. More generally, when $\kp\pm\kc\ne0$, the cell must be smaller than the two-photon wavelength $2\pi/|\kp\pm\kc|$ for motional decoherence to be neglected.

\subsection{Collisional effects}\label{subsec_collisional}
\noindent Up to here, we have discussed the effect of collisions on the atomic motion and the resulting spectral and spatial behaviour. We now turn to consider the effect of the collisions on the atomic internal state. As illustrated in Fig.~\ref{fig:thermal}(d), we consider collisions of active atoms among themselves and collisions with buffer-gas atoms. The latter  predominantly decohere the orbital transitions, leading to so-called pressure broadening $\gamma^\text{col}_{13}$ of the optical line \cite{RevModPhys.54.1103}. This broadening is typically on the order of $\gamma^\text{col}_{13}=10$ MHz per Torr of a buffer gas \cite{Pitz2009}. The cross-section for relaxation and shift of the spin transition (within the ground level) due to buffer-gas collisions is much smaller, and it becomes important, for example, in miniature atomic clocks \cite{vanier_book}. Notably, for ladder EIT which relies on the coherence between electronic orbitals, pressure broadening is always destructive.

On the other hand, collisions amongst the active atoms can introduce substantial spin decoherence $\gamma^\text{col}_{12}$, thus broadening the EIT line in $\Lambda$ systems. There is collisional cross-section for `spin exchange', which conserves the total spin of the colliding pair \cite{HapperPRL1973,KatzPRL2015}, and a cross-section for `spin destruction', which relaxes the total spin. The former is usually larger and can reach the kHz regime. The rate of both processes depends on the collision rate, which is linear in the atomic density. This leads to a trade-off when determining the  density, as discussed in the next section, with higher densities providing for stronger EIT at the expense of faster collisional relaxation and broader lines.

The spectral broadenings due to collisions $\gamma^\text{col}_{13}$ and $\gamma^\text{col}_{12}$ are typically considered to be homogeneous, adding to the natural homogeneous linewidths $\gamma_{13}$ and $\gamma_{12}$. For a $\Lambda$-type EIT, in the limit $\gamma_{13}+\gamma^\text{col}_{13}\gg\sigma_\text{Dop}$ ({\it i.e.}, when the inhomogeneous Doppler broadening of the optical line is relatively small), the EIT linewidth obtains the simple form
\begin{equation}
    \geit=\frac{\Omega_\mathrm{c}^2}{\gamma_{13}+\gamma^\text{col}_{13}+\sigma_\text{Dop}}+\gamma_{12}+\gamma^\text{col}_{12}.
    \label{eq:gammaEITcollisions}
\end{equation}
Equation (\ref{eq:gammaEITcollisions}) also provides a crude approximation when $\gamma_{13}+\gamma^\text{col}_{13}$ and $\sigma_\text{Dop}$ are comparable; For a more exact result, a calculation including an integration over thermal velocity groups is required. In the general case the calculations typically require numerical calculations of the Voight integral; however, by approximating the Maxwell velocity distribution with a Lorentzian lineshape, one can obtain a more precise analytical approximation to the EIT linewidth for the case of an arbitrary ratio between homogeneous and inhomogeneous broadenings of the optical transitions~\cite{lee'03}. 
 
\section{How to design the ``right'' EIT realization in hot vapor}
\label{Section:designEIT}
\subsection{Important parameters}
\begin{figure}[tb]    \centering
    \includegraphics[width=\columnwidth]{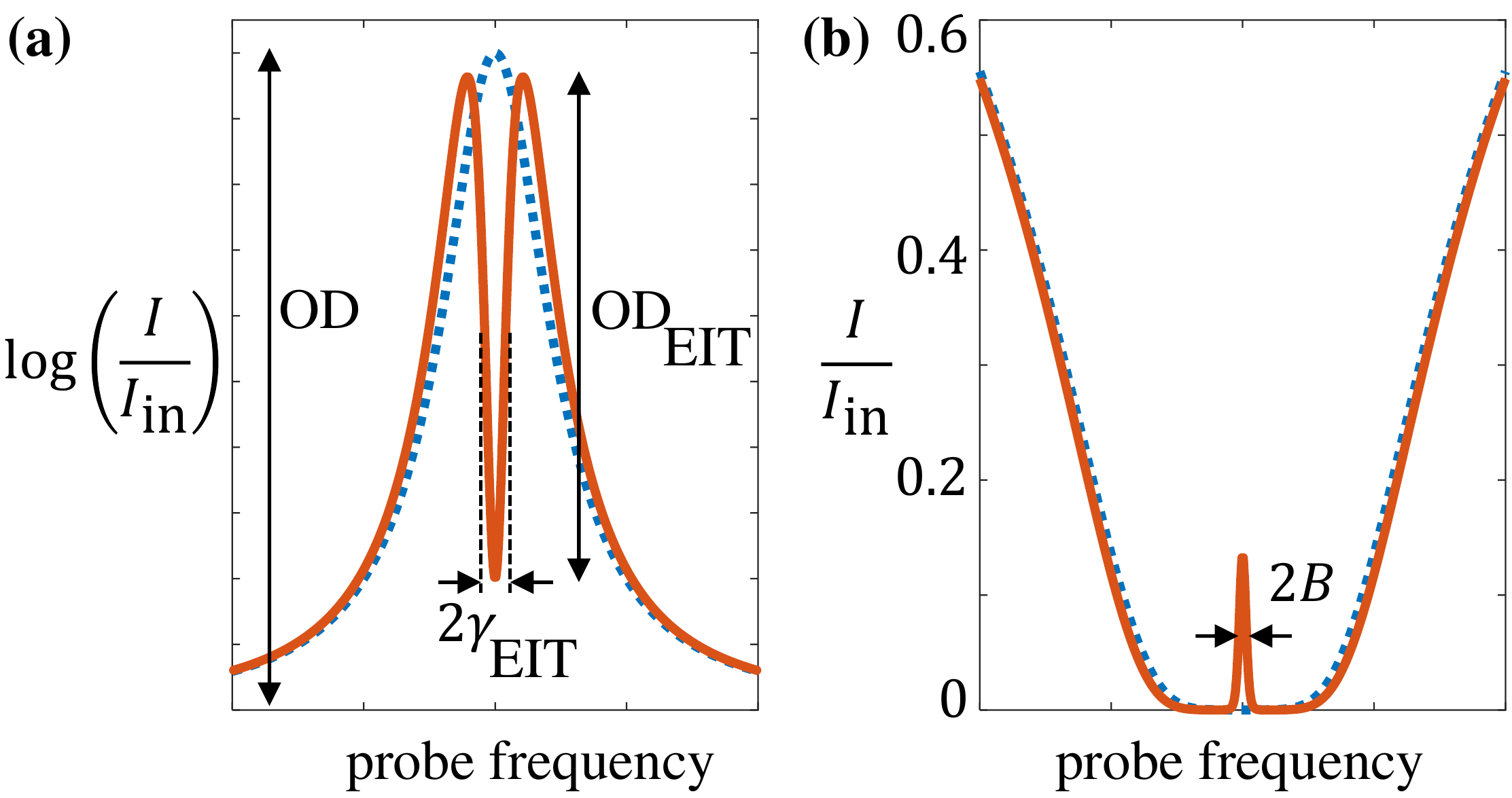}\caption{\textbf{Key parameters for successful EIT applications.} (a) extracting three important parameters, $\OD$, $\ODeit$, and $2\geit$, from the logarithm of the EIT absorption spectrum. (b) the absorption spectrum as measured, illustrating the difference between the bandwidth $2B$ and linewidth [$2\geit$ in (a)].}
    \label{fig:parameters}
\end{figure}

\noindent Different figures of merit of the EIT process may be relevant for different applications. Nevertheless, there are three parameters measurable directly from the EIT absorption spectrum, as illustrated in Fig.~\ref{fig:parameters}(a), that together govern most EIT applications. The first parameter is the resonant optical depth of the medium in the absence of control field. In terms of the absorption coefficient $\alpha_0$, the optical depth is defined by $\OD=2\alpha_0 L$, where $L$ is the medium's length. The other two parameters are: $\ODeit$, the reduction in the optical depth at the EIT resonance (defining the relative height of the transmission line); and $2\geit$, the full width of the EIT transmission line. To the leading order, these parameters determine the transmission bandwidth, the slow-light (group) delay, and the sensing response, as well as many other important properties.

The transmission on the EIT resonance is given by $e^{-\OD+\ODeit}$. High transmission therefore requires that $\ODeit$ approaches $\OD$, which is often referred to as high-contrast EIT.
The frequency range around the resonance for which the transmission is still relatively high (\textit{e.g.}, higher than $e^{-1}$ of the maximal transmission $e^{-\OD+\ODeit}$) is denoted as the bandwidth $B$. The bandwidth determines, for example, how short the (slow-light) probe pulse can be before suffering from significantly increased absorption and distortion.  
Notably, $\geit$ is the half-width of the EIT line after taking the logarithm of the absorption spectrum, and it is equal to $B$ only for small $\ODeit$, see illustration in Fig.~\ref{fig:parameters}(b). As $\ODeit$ increases, the bandwidth decreases, and it is given by $B= \geit/\sqrt{\ODeit-1}$, sometimes referred to as the EIT density narrowing~\cite{lukin'97prl}. Another aspect is the delay of slow light in the medium, which is given by $\tau=\ODeit/\geit$. It follows that the time-bandwidth product, which indicates how many separated probe pulses can fit inside the medium, is given by $B\tau\approx \sqrt{\ODeit}$ for $\ODeit\gg 1$.

More generally, $\OD$ determines the maximal strength of the cooperative light-matter interaction, while $\ODeit$ determines the effective fraction of the $\OD$ that is available for controllably coupling light to the $\rho_{12}$ coherence. Therefore, high-fidelity protocols based on EIT require high $\OD$ and $\ODeit\approx\OD$. For example, the inefficiency of light storage, for probe pulses within the bandwidth $B$, scales as $1/\ODeit$ ~\cite{gorshkovPRL}.

To maximize $\OD$ and $\ODeit$ while minimizing $\geit$, one desires an optically thick medium with slow spin decoherence processes and a strong control field. This desire encounters many practical trade-offs, which we detail in the next section. 

\subsection{Trade-offs}\label{subsection:tradeoffs}
\noindent An ideal implementation of EIT requires simultaneously using all ideal resources: the highest OD possible, the strongest control Rabi frequency, and the longest coherence time. For example, for high bandwidth operation, one needs a strong Rabi frequency, balanced by the corresponding high OD, to maintain high contrast and low group velocity.
However, in real life, all of these resources are limited and can often come at the expense of one another. Designing the experiment is thus the art of striking the balance between different requirements in a way that best serves the goal of a specific experiment. In the following, we list some common trade-offs and note the tensions that may rise when trying to optimize experimental configurations. This is far from being an exhaustive list, and is meant to provide some general guidance while navigating this many-parameter space.

Most EIT experiments use narrowband lasers -- either diode or solid-state lasers -- with output power usually limited to a few Watt. To account for the limited power and maintain high Rabi frequency, a typical solution is to focus the control field down to increase the local intensity. This, however, limits the size of the probe field which must be contained within the control field. These finite fields deviate from the ideal, plane-wave description laid out in Section \ref{Section:concept} and result in both transit-time broadening and inhomogeneity of the control Rabi frequency in the transverse dimension.

To increase the OD, one can increase the temperature of the vapor cell and by that the atomic density. However, this would also increase the collision rate that may reduce spin lifetime and limit the efficiency of optical pumping. This trade-off can be partially relaxed by using spin transitions that are free of spin-exchange relaxation \cite{Walker1997,Katz2018}, but spin destruction is always present. In addition, increasing OD results in the enhancement of non-linear light-matter interactions, such as four-wave mixing~\cite{phillipsPRA11,laukPRA13}. Another route to increasing the OD is to elongate the cell, which can result in a non-compact geometry and, more fundamentally, limits the size of the probe and control fields as paraxial diffraction becomes important.

To prolong the spin coherence lifetime, one can increase buffer-gas pressure and thus slow down diffusion to eliminate motional decoherence. However, this increases pressure broadening and would ultimately introduce spin relaxation via collisions. Plus, since collisional dephasing has a much stronger effect on the excited electronic states, this strategy is not possible in ladder EIT. One can decide to work with wall-coating instead of a buffer-gas, in order to reduce spin relaxation at the walls, however current coatings are temperature limited to about 50-100~$^\circ$C, which limits the atomic density. In case of non-degenerate three-level system, the length of a coated cell must be smaller than the two-photon transition wavelength to avoid spin-wave phase variation across the cell.

\section{How to build basic experimental setups}
\label{section:buildEIT}
\begin{figure}[tb] \centering
\includegraphics[width=\columnwidth]{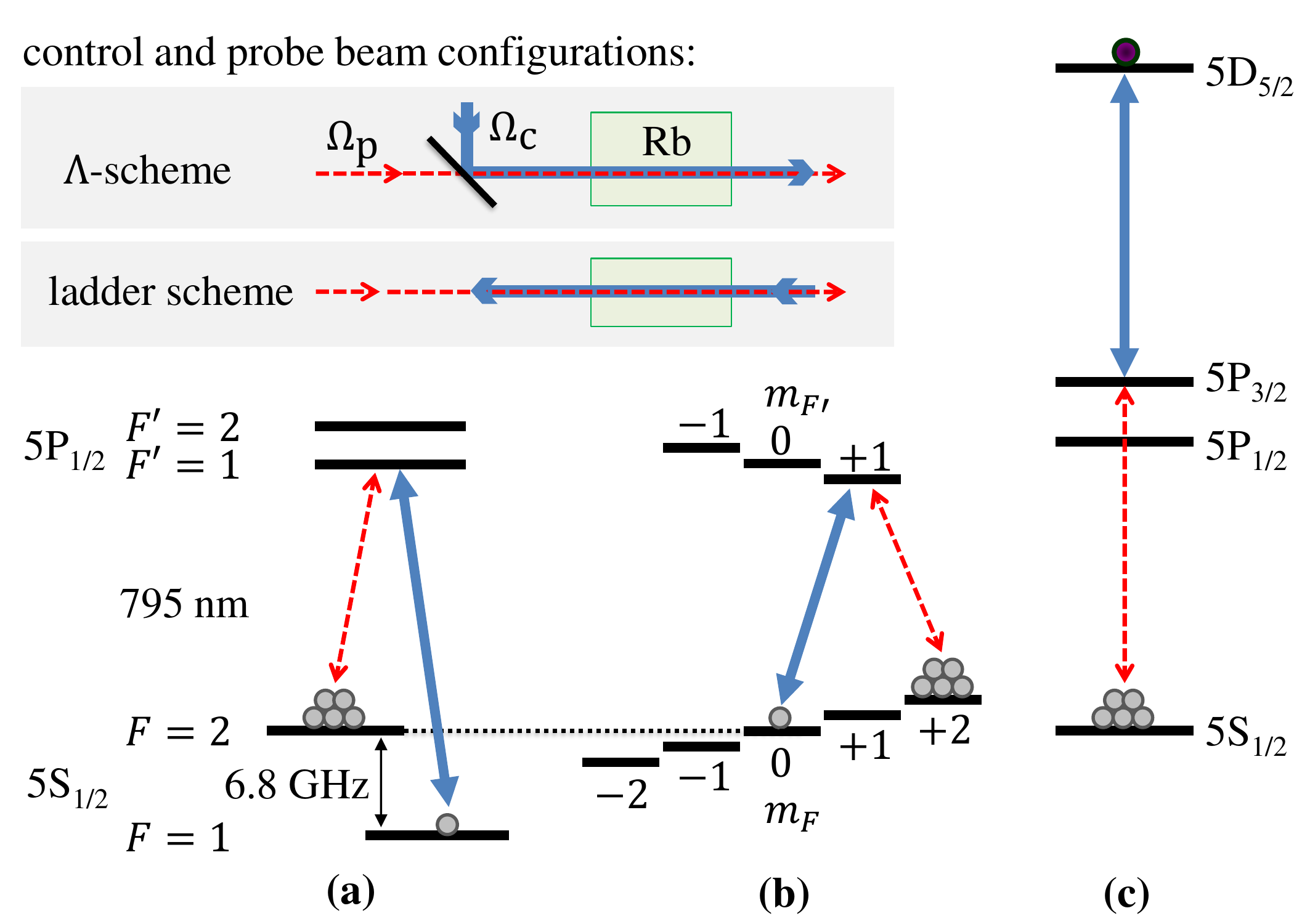}
\caption{\textbf{
Control and probe transitions (solid and dashed arrows, respectively) typical three-level EIT schemes in $^{87}$Rb.} In the examples depicted here, the levels $|1\rangle$, $|2\rangle$, and $|3\rangle$ from Fig.~\ref{fig:LambdaEIT}(a) correspond, respectively, to: (a) the $|F = 2\rangle$ and $|F = 1\rangle$ hyperfine levels of the $|5S_{1/2}\rangle$ ground-state, and the $|F' = 1\rangle$ hyperfine level of the $|5P_{1/2}\rangle$ excited state; this is the ``hyperfine" EIT configuration. (b) The magnetic sub-levels $|m_F = 2\rangle$, $|m_F = 0\rangle$ of the $|5S_{1/2}, F = 2\rangle$ ground-state, and the magnetic sub-level $|m_F' = 1\rangle$ of the $|5P_{1/2}, F' = 1\rangle$ excited state; this is the ``Zeeman" EIT configuration. (c) The states $|5S_{1/2}\rangle$, $|5D_{5/2}\rangle$, and intermediate state $|5P_{3/2}\rangle$; this is the ``ladder" EIT configuration.
   \textbf{Inset:} In hyperfine and Zeeman EIT, the co-propagating arrangement suppresses Doppler broadening, as the control and probe frequencies are very close. In ladder EIT, the  counter-propagating arrangement minimizes Doppler broadening (as the two fields are absorbed during the excitation; see Sec.~\ref{Section:EITinvapor}.A).  
}
 \label{fig:3lvl}
\end{figure}

\noindent A simple three-level atom as described in Sec.~\ref{Section:concept} does not really exist. Nevertheless, the three-level model often well captures the interaction in real atoms, such as in Alkali metals. These atoms offer two-photon ground-state transitions that are accessible with relatively low optical power. The ready availability of inexpensive robust laser diodes at wavelengths corresponding to the D lines in Rb and Cs make these particular alkalis especially favored~\cite{MacAdamAJP1992SatSpec}.
The most widely used three-level EIT schemes with alkali atoms correspond to the three configurations depicted in Fig.~\ref{fig:EITconfigs}. Examples for the implementation of a hyperfine $\Lambda$ scheme, Zeeman $\Lambda$ scheme, and a ladder scheme are illustrated for $^{87}$Rb in Fig.~\ref{fig:3lvl}(a-c). 
Ladder EIT has recently received increased attention in the context of the use of Rydberg atoms for quantum information experiments \cite{2016ofer,2020adams}; With a two-photon excitation, the level $|2\rangle$ of the three-level model can be either a $nS$ or a $nD$ Rydberg state. 

A stable, relative phase between the control and probe fields is required to create the long-lived ground state coherence needed for a robust dark state. 
Zeeman EIT is the least resource-intensive of the three schemes above, requiring only acousto-optic modulation of the output from a single laser at a few tens of kHz to generate the control and probe frequencies resonant with the Zeeman-shifted sub-levels, as depicted in Fig.~\ref{fig:3lvl}(b). Hyperfine EIT requires two phase-locked lasers each tuned to the hyperfine optical transitions, as depicted in Fig.~\ref{fig:3lvl}(a)~\cite{MarinoLaserLock2008,Appel_2009}. Alternatively, one can use various modulation techniques to derive both fields from the same laser.  The choice of modulation technique depends on the application, and on the alkali metal used. If the modulation frequency is smaller than $\approx 3.5$ GHz, one can generate a probe field by double-passing a high-frequency acousto-optical modulator (AOM)~\cite{brimroseAOM}; this approach works well for alkali metals with smaller ground-state  hyperfine splitting values, such as Na and ${}^{85}$Rb~\cite{narducciPRA04,lettPRA08}. The advantage of such an approach is that the generated probe field is a physically separate beam, and its optical properties can be manipulated independently of the control field. The drawback is the low efficiency of AOMs at such high frequency. If higher modulation frequency is required, electro-optical modulators are the best choice~\cite{novikovaPRA08,10.1117/12.2186639}; modern fiber EOMs are broadband and can achieve high modulation efficiency even at moderate rf power. The drawback of this method is that the carrier and all modulation sidebands emerge perfectly spatially overlapped, so that either the experiment must operate with the control and probe fields in the same spatial mode and polarization, or additional filtering must be employed to separate the desired modulation sideband. Moreover, potential complications may arise due to coupling of the off-resonant modulation sideband into the atomic system, for example leading to undesirable multi-wave mixing processes \cite{phillipsPRA11}. Also, while most of the diode lasers do not allow direct current modulation above 1 GHz, vertical cavity surface emitting lasers (VCSELs) have very fast response time  and  can be modulated at frequencies up to tens of GHz~\cite{mikhailov2009AJP_clock_for_undergrads}. This modulation method is particularly attractive for development of chip-scale atomic clocks and magnetometers~\cite{kitchingAPR2018,KnappeCSAC2004,SchwindtAPL04_chipEITmagn,doi:10.1063/1.2838175}. However, the low output power and broad linewidth of VCSELs limits their applicability for most other situations. Finally, ladder-EIT schemes, especially those involving Rydberg atoms, use control and probe lasers that are much farther apart in frequency, by up to hundreds of nanometers; in this case the frequency lock is often used to keep the lasers on the two-photon resonance~\cite{AbelAPL2009_RydbergEITfreqlock}.   

\begin{figure*}[tb] \centering
\includegraphics[width=\textwidth]{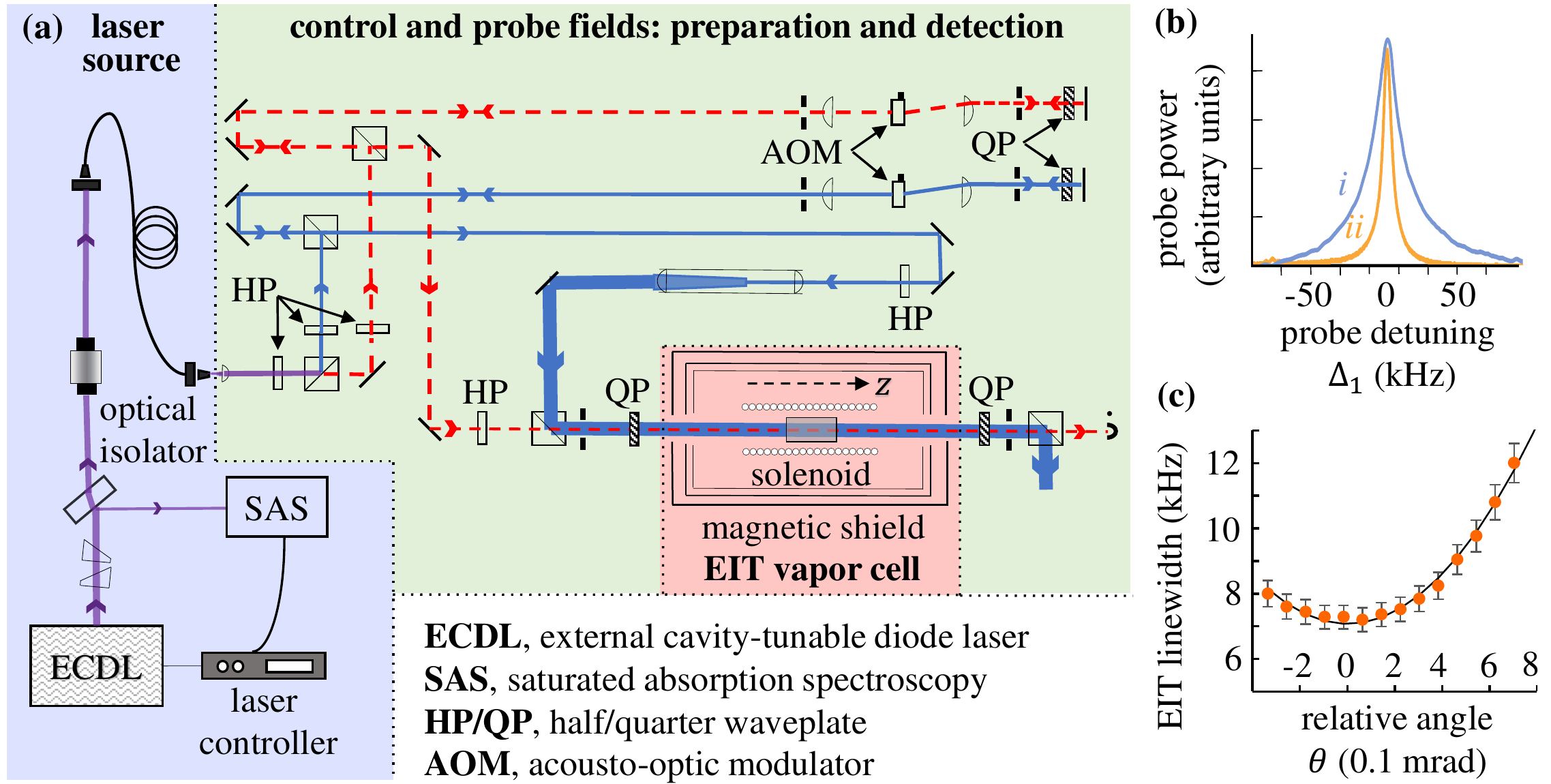}
\caption{\textbf{(a) Optical layout for Zeeman EIT in $^{87}$Rb.} The three main sections, namely, the laser source, the preparation and detection of control and probe fields, and the EIT vapor cell, are delineated. The laser at 795 nm is split at a polarizing beamsplitter into a weak probe beam (dashed line) and a strong control beam (solid line). Each beam is sent through an acousto-optical modulator set up in double-pass mode. The two beams are re-combined and passed through a quarter waveplate to create control and probe beams of mutually orthogonal circular polarization before insertion into the $^{87}$Rb vapor cell. The cell
is enclosed by a solenoid that provides a $B$-field along the laser propagation direction $z$ to create the Zeeman three-level scheme in Fig.~\ref{fig:3lvl}(b), and the setup is magnetically shielded. On exiting the cell, the control beam is reflected away, while the probe transmission is focused onto a fast photodiode. The probe frequency is scanned around the (fixed) control frequency, yielding an EIT spectrum. The control beam is expanded in size to provide a homogeneous Rabi frequency over the transverse extent of the probe.(b) Typical Zeeman EIT spectra (normalized to maximum transmission) observed for (i) high and (ii) low control intensity in a few $\mu$Torr of warm isotopically pure $^{87}$Rb vapor with several Torr of inert buffer gas (see text for details). 
(c) Dicke-narrowed EIT linewidths measured at fixed control intensity as a function of a small relative angle $\theta$ between the control and probe beams. The solid line is a quadratic fit. Linewidth broadening in warm-vapor EIT due to angular misalignment is significantly reduced by the presence of Dicke narrowing.
}
\label{fig:optlayout}
\end{figure*}

An optical layout showing some general features of an EIT setup, in the specific context of Zeeman EIT in $^{87}$Rb, is depicted in Fig.~\ref{fig:optlayout}(a) and briefly discussed below. For further experimental details, please refer to Ref.~\cite{derose2022producing}.

\subsection{Laser source} 
\noindent For many alkali-based EIT experiments, it suffices to use lasers that provide low power ($20-40$~mW), have a linewidth much narrower than the Doppler linewidth in warm vapor ($\ll 100$ MHz), and are frequency-tunable over the ground state hyperfine splitting ($2-10$~GHz, depending on choice of alkali). Grating-feedback-tuned external cavity diode lasers (ECDL) with a typical laser linewidth of a few hundred kHz, or distributed feedback diode lasers (DFB) with linewidth of 1-2MHz, satisfy all these parameters and are commercially available; these lasers are widely used in EIT experiments.
One of the known problems of the diode laser is asymmetry and astigmatism of the output beam. To solve this problem, typically the output of a diode laser is coupled into a single-mode polarization maintaining optical fiber, as shown in Fig.~\ref{fig:optlayout}(a). The main purpose of the fiber is to create a clean Gaussian output beam profile with polarization drift of no more than a few percent. An optical isolator, placed in between, suppresses frequency-destabilizing back-reflection from the fiber into the laser. If the output beam out of a diode laser is strongly anisotropic, an anamorphic prism pair can be used to circularize its elliptical cross-section and improve coupling efficiency into the fiber. Figure \ref{fig:optlayout}(a) also shows that a small portion of the ECDL output is split off with a glass window to stabilize the laser frequency near the $F = 2 \rightarrow F' = 1$ D1 transition [see Fig.~\ref{fig:3lvl}(b)] using, for example, the method of saturated absorption spectroscopy (SAS)~\cite{MacAdamAJP1992SatSpec}.

Solid state laser systems, such as stabilized continuous-wave (CW) Ti:Sapph lasers~\cite{MSquaredLaser} are the best for an experiment requiring higher laser power, as they can output up to several watt of narrow-band optical field, tunable in a very wide spectral range. These laser systems also provide a much cleaner output spatial mode and do not have a broad-spectral background emission that is typical of diode lasers. However, they are significantly more expensive.

\subsection{Control and probe field preparation and detection}
\noindent The output from the fiber is split into two orthogonally linearly polarized beams with a polarizing beamsplitter (PBS) to create a weak probe beam (dashed line) and a strong control beam (solid line). Each beam is fed into an acousto-optic modulator (AOM) to enable tuning of the control and probe frequencies so as to satisfy the two-photon resonance condition. The AOMs also enable the scanning of the probe frequency around the fixed control frequency, where the double-pass AOM configuration suppresses spatial misalignment of the beam during the scanning. To avoid slow frequency and phase drifts between the control and probe fields, the AOMs are driven by twin phase-locked rf signals from the same dual-output waveform generator. 
   
As described in Secs.~\ref{Section:concept} and \ref{Section:EITinvapor}, the linewidth of the EIT resonance $\geit$ depends upon the control field intensity, which should therefore be ideally uniform in the region of spatial overlap with the probe. One method to approximately satisfy this is to expand the control beam to a size significantly larger than the probe, prior to recombining the two beams and inserting into the vapor cell. Expanding the control beam also allows us to use a larger probe beam, which helps combat transit-time broadening in vapor cells without buffer gas or wall coatings, as indicated previously in Sec.~\ref{Section:designEIT}B. 

For Zeeman EIT, the control and probe beams must have mutually orthogonal circular polarization, which is achieved by placing a quarter waveplate (QP) in the path immediately before the cell. Upon exiting the cell, a second QP converts the circular polarizations back into linear, which enables separation of the probe and control beams at a PBS. The weak probe transmits through and is focused onto a photodiode, while the strong control beam is reflected away. It is important to minimize leakage of the strong control beam into the photodiode; a Glan-Taylor prism with an extinction ratio of $10^5$ may be employed for the probe-control beam separation [all other PBS in Fig.~\ref{fig:optlayout}(a) provide a typical extinction of $10^3$]. 

\subsection{EIT vapor cell} 
\noindent A typical atomic vapor cell is a pyrex cylinder (or sometimes a cube), with a stem (side arm) containing a few mg of solid Rb or Cs. If the cell has bare polished glass windows, one must expect approximately $20\%$ losses due to reflection of the windows. Moreover, the reflected beams may perturb the atomic coherence and distort the EIT signal~\cite{MatskoPRA2007retroreflectionEIT}. Luckily, vapor cells with anti-reflection coatings on both sides of the quartz windows are now available.  

The content of an experimental cell also depends on the details of the applications. One ingredient -- alkali metal -- is necessary, of course. The choice of specific element is sometimes determined by its properties, such as the number of Zeeman sublevels or the separation between excited state levels, but often is driven by the available equipment and previous work of the research group. Also, for alkali metals having more than one isotope, it is important to decide if natural abundance mixture of isotopes is sufficient, or isotopically enriched metal must be used (at an additional cost). 

The next step is to choose any additions to the pure metal, especially for the experiments with $\Lambda$-type interactions. As discussed in Sec.~\ref{subsec_diffusion}, in many experiments a few Torr of inert ``buffer" gas, such as He, Ne, or N$_2$, are introduced to extend the confinement time of alkali atoms within the illuminated volume by orders of magnitude~\cite{wynands'99}. 
%
Another possible approach is to coat the cell walls with a derivative of paraffin (tetracontane C$_{40}$H$_{82}$ seems to be the top choice), to help preserve spin coherence during atom-wall collisions. 
Paraffin coatings break down at higher temperatures (typically 60-80~$^\circ$C)~\cite{SeltzerJCP2010coatings}. If high atomic density is required, one may opt for OTS (octadecyltrichlorosilane) coatings which withstand higher temperatures~\cite{SeltzerJAP2009OTScoating}, although the spin-preserving characteristics of OTS seem to be inferior compared to paraffin. Also, since very few commercial sources for wall-coated cells exist, it may not be trivial to procure an antirelaxation coating cell, causing some users to just develop their own in-house coating capabilities.

To control the density of atoms in the experiment, the stem containing the metal can be slightly heated, changing the saturated pressure of alkali vapor, typically   on the order of a few $\mu$Torr, corresponding to an atom number density of $10^{11}-10^{12}$ cm$^{-3}$~\cite{Stecknumbers}.
It is important to heat the cell uniformly and use a bifilar heating wire or other heating methods which do not inject undesired magnetic fields into the sample. The heating wire may be wrapped around the innermost magnetic shield layer (see below) to allow air to circulate for uniform heating. 

The optimal operational temperature is typically determined by optimizing the experimental performance. On one hand, as indicated in Sec.~\ref{Section:designEIT}A, the EIT linewidth $\geit$ is modified by the optical depth of the sample to yield a useful
``effective EIT bandwidth" given by $B \sim {\geit / \sqrt{\ODeit}}$ (see Fig.~\ref{fig:parameters}). Raising the atom density raises the optical depth and narrows the bandwidth $B$. 
Moreover, Eqs.~(\ref{gammaEIT}) and (\ref{vgroup_EIT}) explicitly show that the group velocity $v_\text{g}$ is lowered at higher atom density $N$. 
Clearly, keeping the number density $N$ high is desirable for many EIT-based quantum applications. On the other hand, raising $N$ too much may actually deteriorate EIT performance, by, \textit{e.g.}, increasing undesirable spin-relaxation collisions (see Sec.~\ref{subsection:tradeoffs}) and introducing additional density-dependent relaxation mechanisms. For example, radiation trapping -- the reabsorption of spontaneously emitted photons  within the illuminated volume --  reduces the average spin coherence lifetime. It becomes significant at densities $\geq 5 \times 10^{11}$/cm$^3$ for the typical mm-sized laser beam diameters that are employed in vapor cell experiments~\cite{2001matsko}.

Another important decoherence mechanism that contributes to the ground-state decoherence rate is residual magnetic field inhomogeneity~\cite{XiaoEITwidthreview2009,derose2022producing}. In the case of Zeeman EIT, for example, this causes spatial variation of the dark state leading to dephasing. As indicated in Fig.~\ref{fig:optlayout}(a), the cell is placed inside a solenoid which provides controlled dc magnetic fields along the laser propagation direction. The solenoid can be used either for near-complete cancellation of residual fields, so that the Zeeman sub-levels stay degenerate, or for adding a well-defined magnetic field (typically 10-100 mG) that separates the Zeeman sub-levels. This separation is advantageous for both Zeeman and hyperfine EIT when only one desired transition (\textit{e.g.}, the clock transition $m_F=0$) is to be addressed. 
The vapor cell and solenoid are enclosed by a triple-layer of high-magnetic permeability alloy which shields the cell from unwanted external magnetic fields. Before taking any data, the shield must be demagnetized by a degaussing technique using coils wound around one or more of the shielding layers~\cite{derose2022producing}.

\subsection{Probe transmission spectra} 
\noindent To generate an EIT spectrum, the two-photon detuning is adjusted by varying the relative detuning between the probe and control frequencies, and the probe transmission is recorded. This is achieved by scanning the probe AOM around the control frequency which is held fixed. In the Zeeman EIT scheme depicted in Fig.~\ref{fig:3lvl}(b), the Zeeman shifts between the magnetic sub-levels of the $F = 2$ ground state are 0.7 kHz/mG, so a $B_z$ field of 50 mG from the solenoid splits adjacent ground sub-states 35 kHz apart, and a probe scan of less than $\pm 100$ kHz suffices to record the EIT spectral feature. In Fig.~\ref{fig:optlayout}(b), Zeeman EIT spectra observed in a few $\mu$Torr of isotopically pure $^{87}$Rb vapor with 10 Torr of Ne buffer gas at about 60~$^\circ$C are shown for two different control intensities: (i) 5.5 mW/cm$^2$ and (ii) 1.3 mW/cm$^2$, corresponding to $\Omega_\mathrm{c} / 2\pi = 3.9$ MHz and 1.9 MHz, respectively. The spectra are displayed on the same vertical scale to enable a visual linewidth comparison. The FWHM, extracted by Lorentzian fits that are normalized to the same amplitude at zero detuning, yield EIT linewidths of 26 kHz and 7.3 kHz respectively. The ratio of almost 4 between the two linewidths, as also between the two control intensities, is indeed what one expects if the EIT lines are dominated by power broadening. 

Perfect co-linear alignment of the control and probe beams in EIT is an idealization. In the experiment, a slight angle between the two beams exists, which introduces a residual or two-photon Doppler broadening, as discussed above in Sec.~\ref{Section:EITinvapor}A. For the $\Lambda$-scheme with close-lying control and probe frequencies, we may straightforwardly estimate the residual Doppler broadening to be $\keff \cdot \vec{v} \approx |\vec{k} \, |v_\text{th} \, \theta = \sigma_\text{Dop} \, \theta$. Thus, even for a slight relative angle $\theta \sim 0.1$ mrad, with typical Doppler broadening of a few hundred MHz in the warm vapor cell, we see that the two-photon Doppler broadening is tens of kHz and is significant in $\Lambda$-type EIT. However, in the presence of a buffer gas, frequent velocity-changing collisions prevail, which cause Dicke narrowing of the Doppler broadening as discussed in Sec.~\ref{Section:EITinvapor}B. It is mentioned there that the resulting Dicke-narrowed EIT linewidth follows a quadratic dependence on $\theta$ instead of linear; this statement is borne out by the data in Fig.~\ref{fig:optlayout}(c). Here, the control and probe beam sizes ($1/e^2$-radii 2.55 mm and 0.47 mm, respectively) and intensities (1.3 mW/cm$^2$ and 0.12 mW/cm$^2$, respectively) are the same as in curve (ii) of Fig.~\ref{fig:optlayout}(b). The optical depth is $\sim 10$. The solid line is a quadratic fit to the data. The data shows that the EIT linewidth remains narrow despite angular mismatch between the control and probe beams. Thus the tolerance for angular deviation in applications utilizing warm vapor is increased significantly owing to the presence of Dicke narrowing~\cite{shukerPRA2008}. 

\section{Conclusion and outlook}
\label{Section:conclusions}
\noindent Atomic vapor cells bear with them the promise of a feasible and scalable quantum technology. In the last two decades EIT and its analogues have been considered for a wide range of applications, some of which have already matured into practical devices. Below we briefly discuss some of the notable ones enabled by EIT in atomic vapor.

\textit{EIT-based metrology tools.---}
Measurements of energy separation between various alkali spin states lie at the heart of many precise atomic clocks and magnetometer~\cite{vanier_book}. The frequencies of these transitions are in the radio-frequency or microwave spectral range, which hinders the development of compact devices as their size may be limited by the transition wavelength. Since a $\Lambda$-system EIT allows all-optical coupling to these transitions, it is particularly attractive for miniaturization purposes~\cite{vanier05apb}. In particular, there has been a lot of progress on the development of the chip-scale atomic clocks (CSAC) that use microfabricated vapor cells and now achieve fractional frequency stability down to $10^-11$~\cite{Cash_CSACReview_IEEE2018}. The development of EIT-based magnetometers has attracted less attention~\cite{scully92prl,stahler01eurlett,SchwindtAPL04_chipEITmagn}, although several publications have pointed out some of their unique capabilities, \textit{e.g.}, the ability to determine the direction of the magnetic field~\cite{yudinPRA2010,mikhailov2010compass}. 

It is more common to see references to CPT-based atomic clocks and magnetometers, since the majority of experiments use a balanced $\Lambda$-system with equal or nearly-equal intensities of the two optical fields. This configuration allows reduction of some systematic effects like light shifts and can be realized experimentally via direct frequency modulation of a single laser. VCSELs are often used thanks to their excellent modulation response in the GHz range~\cite{mikhailov2009AJP_clock_for_undergrads}. The miniaturization requirements also lead to the frequent use of vapor cells with relatively high buffer gas pressures. Occasionally, a mixture of gasses is used to reduce the temperature variation of the collisional shifts and improve clock stability~\cite{DengAPL2008}. Also, a dynamic Raman-Ramsey interrogation scheme has been successfully implemented to suppress the power broadening of EIT/CPT resonances~\cite{ZanonPRL05,ZanonPRA2011,Liu:13}. 

More recently, ladder EIT resonances have found use for electrometry that takes advantage of the extreme sensitivity of the Rydberg atomic state to external electric fields~\cite{mohapatra_giant_2008}. Due to the long lifetime between an electronic ground state and a Rydberg state it is possible to obtain relatively narrow transmission resonances even in a thermal vapor~\cite{Mohapatra2007,kubler_coherent_2010}. Monitoring the transmission of the ``bottom'' optical field of the ladder [$\mathrm{E}_1$ in Fig.~\ref{fig:EITconfigs}(c)], one can accurately track the energy splitting and shift of the Rydberg level, thus gaining information about external microwave or rf electric field~\cite{sedlacek_microwave_2012,gordonAIP2019,JauPRAppl2020,FancherIEEE2021}. The use of  vapor-cell EIT-based Rydberg atomic systems has been also explored as an alternative technology for audio and video receivers~\cite{CoxPRL2018,prajapati2022arxiv}, spectrum analyzer~\cite{MeyerPRappl2021}, SI-traceable standards~\cite{HollowayJAP2017}, etc.

\textit{Quantum information processing.---}
As we have mentioned in Section \ref{Section:concept}D, the fact that EIT is intimately linked to the existence of a dark state and permits dynamic control over the group velocity of the dark-state polaritons can be utilized to store photonic information. The information could be encoded by a number-state qubit (for example zero excitation or a single excitation in a mode), a polarization qubit, a time-bin qubit, or a spatial dual-rail qubit. Such information can also be encoded in continuous variables, as the quadratures of the electromagnetic field. EIT based memories, both in hot vapor as well as in ultracold atoms, were shown to be tuned to reach very high efficiency, noise-free operation, and large bandwidth \cite{TaipeiPRL2018,Finkelstein2018}. Ongoing efforts aim to combine several of these and  first commercial applications are already available \cite{qunnectSPIE2022}. The material part of such dark-state polaritons can also be realized as a Rydberg excitation. Working at the single-photon level, strongly interacting Rydberg dark-state polaritons have been shown to mediate effective photon-photon interactions resulting in phenomena such as single photon trains, photonic molecules, single-photon transistor, and controlled phase gate between photons \cite{Vuletic2013b,HofferberthPRL2014,2016ofer,LahadPRL2017,tiarks2019photon}.   

\textit{EIT as a spectral filter.---}
Narrow EIT resonances can be used to shape the fluctuations of broadband optical probe field, as only the frequencies within the EIT transmission window will emerge after the interaction. The narrow (sub-MHz) tunable EIT linewidth provides a potential alternative to traditional spectral filters based on optical cavities, as both transmissive and dispersive properties of EIT resonance can be adjusted via the control field parameter~\cite{mikhailovPRA2006_EITfilter,Keaveney_2014}.
In particular, several experiments demonstrated the transmission of squeezed light through EIT~\cite{akamatsu2004prl,lvovsky09njp_squeeziong_eit}, as well as squeezed pulse propagation in  slow~\cite{akamatsu_ultraslow_2007,lam2008oe_sq_eit}, stored~\cite{lvovskyPRL08}, and fast light~\cite{mikhailovOL2013fast_squeezing} regimes.

If placed inside a cavity, EIT can serve as a control tool for cavity characteristics. For example, it was recognized relatively early in EIT history that intracavity EIT leads to significant cavity linewidth narrowing~\cite{lukin98opl_eit_cavity,xiao2000opl_eit_cavity}. More recently, the role of intracavity EIT or similar narrow multiphoton resonances have been considered: For example, a proposal for enhancement of the laser gyroscope performance using superluminal intracavity atomic medium has attracted a lot of attention~\cite{shahriar2007pra_fast_gyro}. At the same time, a subluminal regime was shown to lead to reduced sensitivity to cavity fluctuations and correspondingly a more stable laser operation~\cite{PhysRevA.100.023846}. 

The future holds great promise for EIT in atomic vapors. There is a constant development of new paradigms, one recent example we have discussed above is electric field sensors based on EIT with warm Rydberg atoms \cite{Schmittberger2021,superhet2020}. Such developments are accompanied by technological maturation of miniature-cell fabrication \cite{Shkel2017,Holloway2018,csem2014,Shaffer2021}, which provide more robust, versatile, and compact platforms for EIT utilization. In parallel to academic research, the technology is also developed in the industry, in both large and small companies. All this progress suggests that we are far from hearing the last word from this diverse field.

\bibliographystyle{unsrt}
\bibliography{bibliographyRF.bib,bibliographyIN.bib,bibliographySB.bib}
\end{document}